\documentclass{article}

\usepackage{arxiv}

\usepackage[utf8]{inputenc} 
\usepackage[T1]{fontenc}    
\usepackage{hyperref}       
\usepackage{url}            
\usepackage{booktabs}       
\usepackage{amsfonts}       
\usepackage{nicefrac}       
\usepackage{microtype}      
\usepackage{amsmath}
\usepackage{cleveref}       
\usepackage{lipsum}         
\usepackage{graphicx}
\usepackage{doi}
\usepackage{ltablex}
\usepackage{xcolor}

\definecolor{darkgreen}{rgb}{0.0, 0.5, 0.0}

\definecolor{brickred}{rgb}{0.8, 0.25, 0.33}

\usepackage{authblk}

\newbox{\orcid}\sbox{\orcid}{\includegraphics[scale=0.06]{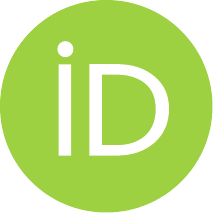}} 

\author[1,2,3]{%
	\href{https://orcid.org/0009-0004-3971-8025}{\usebox{\orcid}\hspace{1mm}Alessandro Bellina\thanks{\texttt{alessandro.bellina@cref.it}}}%
}
\author[4]{%
	\href{https://orcid.org/0000-0003-3221-5973}{\usebox{\orcid}\hspace{1mm}Vito D.~P.~Servedio}%
}
\affil[1]{Centro Ricerche Enrico Fermi, Piazza del Viminale, 1, I-00184 Rome, Italy}
\affil[2]{Sony Computer Science Laboratories - Rome, Joint Initiative CREF-SONY, Centro Ricerche Enrico Fermi, Via Panisperna 89/A, 00184, Rome, Italy}
\affil[3]{Sapienza University of Rome, Physics Dept., P.le A. Moro, 2, I-00185 Rome, Italy}
\affil[4]{Complexity Science Hub, Metternichgasse 8, 1030 Vienna, Austria}

\renewcommand{\shorttitle}{Cognitive Limits Shape Language Statistics}

\hypersetup{
pdftitle={Cognitive Limits Shape Language Statistics},
pdfsubject={},
pdfauthor={Alessandro Bellina, Vito D.~P.~Servedio},
pdfkeywords={},
}

\begin{document}

\title{Cognitive Limits Shape Language Statistics}
\maketitle
\date{today}

\begin{abstract}
Statistical regularities in human language have fascinated researchers for decades, suggesting deep underlying principles governing its evolution and information structuring for efficient communication. 
While Zipf’s Law describes the frequency-rank distribution of words, deviations from this pattern—particularly for less frequent words—challenge the notion of an entirely optimized communication system.
Here, we present a theoretical framework that integrates concepts from information theory, network science, and adjacent possible to explain these deviations. 
We propose that language evolves through optimization processes constrained by the finite cognitive capacities of humans.
This results in a dual structure within language: while frequent words form an optimized, highly navigable core, less frequent words reside in a suboptimal regime, requiring more complex combinations to convey meaning. 
Our findings reveal that Zipf's exponents' deviation to larger values—from 1 to 2—marks a transition from an optimal to a suboptimal state, dictated by cognitive limits. 
This transition imposes a fundamental limit on communication efficiency, where cognitive constraints lead to a reliance on combinations of words rather than the creation of new vocabulary to express an open-ended conceptual space. 
A simple model based on the adjacent possible remarkably aligns with the empirical frequency-rank distribution of words in a language. 
These insights have significant implications for natural language processing and the design of artificial linguistic models, offering new perspectives on optimizing human and machine communication.
\end{abstract}


\section{Introduction}
\label{sec:intro}
    Human language, with its complex structures and dynamics, has been the subject of extensive scientific research~\cite{zipf2013psycho, mandelbrot1961theory}. Over the years, researchers have identified intriguing statistical properties that govern language usage~\cite{zipf2016human, mandelbrot1953informational, baronchelli2012language, naranan1998models}. One of the most significant regularity is Zipf's Law~\cite{zipf2013psycho, de2021dynamical, li2002zipf, i2007universality}, which states that the frequency $f(R)$ of the $R$-th most frequent word is inversely proportional to its rank $R$, following $f(R) \sim R^{-1}$~\cite{newman2005power}. This relation, formulated by George Kingsley Zipf in 1936, can be interpreted as the result of the Principle of Least Effort applied to communication~\cite{zipf2013psycho, zipf2016human, mandelbrot1965information, kanwal2017zipf, zhu2018principle}. In essence, Zipf's Law reflects an optimization process in communication~\cite{ferrer2016compression, ferrer2022optimal, i2005decoding, cubero2019statistical, ferrer2015compression}, interpreted as minimizing the effort between the speaker and the hearer~\cite{ ferrer2018origins, cancho2003least, vogt2004minimum, gerlach2014scaling}. However, the availability of Big Data has allowed for larger-scale analyses, revealing significant deviations from Zipf's Law in the tail of the word distribution~\cite{montemurro2001beyond, li2010fitting}.
    Several studies have recognized that Zipf's Law primarily applies to the most frequent words, often referred to as the \textit{kernel} or \textit{core} of the language~\cite{ferrer2001two, cancho2001small, gerlach2013stochastic, Yu2018zipf}. For infrequent words, which lie in the tail of the distribution, the scaling relation changes to $f(R) \sim R^{-\alpha}$ with $\alpha \approx 2$. 
    
    Recent research have explored the emergence of statistical criticalities as a fundamental characteristic of maximally informative samples~\cite{cubero2019statistical, haimovici2015criticality, marsili2013sampling}, a concept that resonates with the theory of effort minimization in communication. In this context, language evolution can be seen as an optimization process aimed at conveying the maximum possible information about an unknown generative process, and Zipf's Law naturally emerges as the footprint of maximal optimization~\cite{cubero2018minimum}.
    
    Another perspective to uncover the statistical properties of natural language is network representation~\cite{cancho2001small, choudhury2010global, lynn2022emergent, markosova2006language}. In particular, Word Co-occurrence or Word Adjacency Networks consider words as nodes, with links between any adjacent words in the text. A striking property of these networks is that the \textit{kernel} or \textit{core}, consisting of the most frequent words, reveals a topology reminiscent of a Barabási-Albert model~\cite{cancho2001small}. This network structure offers high navigability due to its scale-free properties~\cite{barabasi1999emergence}, which align with the optimization required for efficient communication~\cite{ferrer2022optimal}. In contrast, nodes representing infrequent words still exhibit scale-free properties, but with an anomalous degree exponent~\cite{barabasi2003scale}. These networks lack the optimized organization needed to enhance navigability, which characterizes the kernel~\cite{seyed2006scale}.
    
    In this paper, we propose a theoretical framework to understand the evolution of human language by integrating information theory~\cite{cover1999elements, shannon1948mathematical}, complex network theory~\cite{choudhury2010global} and innovation dynamics concepts~\cite{tria2014dynamics}. Our main hypothesis is that language tends toward optimization but is constrained by finite cognitive capacities, leading to a suboptimal regime for infrequent words. We argue that humans optimize communication for a limited number of frequently used words, while performance decreases as the lexicon expands. This loss of optimization manifests itself in a slowdown in vocabulary growth~\cite{petersen2012languages}, which forces the need to associate meanings with increasingly complex combinations of words, thus extending the length of the text.
    
    Additionally, our approach reveals structural limits on non-optimization that still preserve communication capabilities. This explains why the exponent $\alpha \approx 2$ of the frequency-rank distribution serves as an intrinsic limit in linguistic systems. Finally, we introduce a generative model based on the concept of the Adjacent Possible~\cite{koppl2021explaining, tria2014dynamics, bellina2024time, di2025dynamics, loreto2016dynamics}, which reproduces the evolution mechanism and provides a quantitative account of the observed scaling exponents.

\section{Zipf's Law and optimal compression}

    \begin{figure*}[t]
        \centering
        \includegraphics[width=1\textwidth, trim = 30 30 30 30, clip]{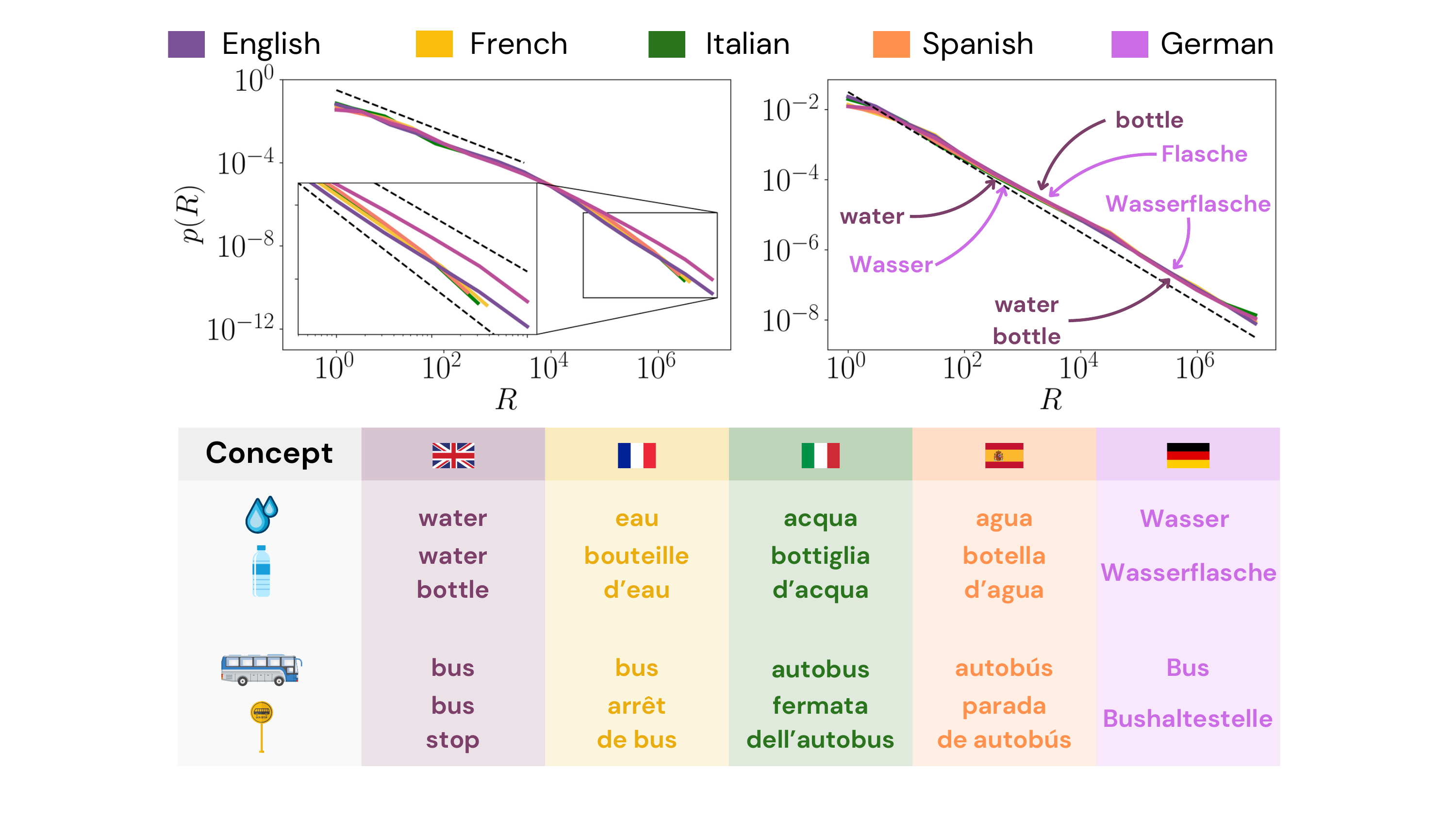}
        \caption{\textbf{Statistical properties of languages.} \textbf{Top left:} Normalized frequency-rank distribution $p(R)$ of single words (1-grams) for English, French, Italian, Spanish, and German. For each of these languages, the most frequent words follow a power law with exponent -1, known as Zipf's Law. In the tail of the distribution, the exponent shifts to 2 for most languages, indicating a transition from an optimal to a suboptimal regime. This shift reflects the need to form combinations of words to represent an increasingly complex conceptual space using a finite vocabulary. In German, however, the exponent bends less sharply, with a value of approximately 1.5. This reflects its strategy of expanding the vocabulary by creating new words from combinations of existing ones~\cite{fleischer2012word}, as shown in the bottom panel. \textbf{Top right:} Frequency-rank distribution of n-grams (obtained aggregating from 1-grams to 5-grams) for the five languages. Here, n-grams are treated as representations of individual concepts, and the resulting distribution follows Zipf’s Law without deviation, indicating efficient concept representation through word combinations. Notably, similar words in related languages appear at comparable positions in the distribution. For instance, in German, the compound word ``Wasserflasche'' appears close in rank to the corresponding English bigram ``water bottle''. This suggests that, despite differences in word formation strategies, these distributions directly reflect underlying conceptual representations. Plots are generated using n-gram frequencies from the Google Ngram database~\cite{michel2011quantitative}. \textbf{Bottom:} In this panel, we illustrate the effect of using word combinations to represent complex concepts. Cognitive constraints limit the indefinite growth of vocabulary, leading to an increased reliance on word combinations as concepts grow more complex. This change impacts the word frequency distribution, causing a shift from an exponent of 1 to 2. German differs since it often creates compound words for complex concepts, thus retaining a lower exponent of $\approx$ 1.5. However, as words can be arbitrarily long as often happens in German, a larger vocabulary does not necessarily imply using fewer characters.}
        \label{fig:fig1}
    \end{figure*}

    Zipf's Law is one of the most fundamental observations in quantitative linguistics~\cite{zipf2013psycho, zipf2016human}, and its origin has been widely debated~\cite{mandelbrot1961theory, i2005decoding, marsili2022quantifying, baek2011zipf, naranan1998models, corominas2011emergence,hanel2020role}. It highlights an essential characteristic of natural languages: a few words are used frequently, while the vast majority are rarely used. Despite its robustness, significant deviations are observed in the tail of the frequency-rank distribution~\cite{ferrer2001two, montemurro2001beyond}. Large corpora of text demonstrate that the exponent of the frequency-rank of words, $-\alpha$, shifts from $-1$ to approximately $-2$ after $R \simeq 8000$. This pattern appears across many languages, as shown in Fig.~\ref{fig:fig1}.
    
    Zipf's Law has often been explained as a consequence of the \textit{principle of least effort}~\cite{zipf2013psycho, cancho2003least, kanwal2017zipf}, which suggests that systems evolve in a way that minimizes some effort or cost~\cite{chaikin1995principles, zhu2018principle}. This principle, which finds applications in various fields such as physics, economics, biology, and psychology~\cite{newman2005power}, implies that language must balance simplicity and expressiveness. This idea has been formalized within information and communication theory~\cite{agouzal2024relationship, i2005decoding, ferrer2022optimal}. Essentially, language evolution can be seen as an optimization process where speakers and listeners minimize their respective efforts in communication while maximizing understanding~\cite{cancho2003least, vogt2004minimum, marsili2022quantifying, cubero2018minimum}.
    
    The trade-off between speaker and hearer efforts can be understood by defining the quantities that each aims to optimize. The speaker's goal is to minimize the effort required to send a message, which leads to a preference for shorter messages. This effort can be quantified by the number of bits necessary to encode the vocabulary (defined as the set of distinct words they use), known as the average code length or resolution~\cite{marsili2022quantifying, cubero2018minimum}. The resolution corresponds to the Shannon entropy of the words, $H(w)$. On the other hand, following~\cite{cubero2018minimum}, the hearer seeks to maximize the information conveyed in the message, which can be evaluated through its relevance~\cite{marsili2013sampling}. Relevance is defined as the Shannon entropy of the word frequencies, $H(f)$.

    As the speaker compresses the vocabulary, it is essential to retain enough information for the hearer to understand the message accurately. This balancing act reaches an optimal solution when the words are distributed according to Zipf's Law~\cite{cubero2019statistical}. In Appendix~\ref{app:A}, we provide the details on the mathematical formulation of the relevance-resolution trade-off and its application to linguistic corpora. 
    
    Thus, Zipf's Law with an exponent of $\alpha=1$ reflects the optimal balance in the speaker-hearer communication process, where the message is maximally compressed without losing information~\cite{cubero2019statistical}. However, this optimization cannot continue indefinitely, as the space of concepts that must be represented is virtually unlimited. Cognitive limitations of the human brain prevent the vocabulary from growing indefinitely~\cite{miller1963finitary, petersen2012languages}, resulting in the shift of the frequency-rank exponent $\alpha$ from $1$ to values larger than $1$. When $\alpha>1$, we enter a regime of lossy compression~\cite{cover1999elements, cubero2019statistical}, where the vocabulary becomes insufficient to represent all the underlying concepts efficiently. 
    
    The only way to cover an unlimited space of concepts with a finite vocabulary is to use combinations of words to represent increasingly complex meanings~\cite{goldberg1995constructions, croft2001radical}. This happens in all the natural languages, as illustrated in Fig.~\ref{fig:fig1}. Let's consider each combination of one or more words as representing a unique concept. The frequency rank distribution of n-grams should follow Zipf's Law, indicating an efficient representation of concepts~\cite{ryland2015zipf, ha2009extending}. In Fig.~\ref{fig:fig1}, we show the frequency rank distribution of n-grams for different languages, which strictly follows Zipf's Law in each case. These observations suggest a direct strategy to enhance language efficiency: identifying the most frequent n-grams and replacing combinations of more than one word with a single codeword. However, this would require a virtually infinite vocabulary, which is prevented by cognitive limitations in the human case. This argument also explains why German has not yet reached the regime exponent of $2$ typical of other languages, and exhibits a smaller exponent of $1.5$ in its tail. German associates unique words with complex concepts, often relying on explicit concatenation of many words~\cite{fleischer2012word}. This reduces the need to rely on multi-word combinations, and thus slows down the exponent from shifting to $2$, keeping it around a smaller value.

    As vocabulary growth slows, it is essential to maintain the ability to create new combinations of words to express new concepts. This implies that there is a lower limit to this slowdown:
    the lexicon must continue expanding at a sufficient rate relative to the text length to enable the formation of new word combinations. If this expansion becomes too slow, there will not be enough words to generate the necessary combinations for representing an open-ended conceptual space effectively. Vocabulary growth follows Heaps' Law, another fundamental property of human language, which states that the dictionary size increases sublinearly as a power law of the text length, i.e. $D(t) \sim t^{\beta}$~\cite{heaps1978information, serrano2009modeling, bernhardsson2009meta, eliazar2011growth}. Moreover, the Heaps' exponent is related to Zipf's exponent as $\alpha = 1 / \beta$~\cite{van2005formal, de2021dynamical}. The analysis of the Word Co-occurrence Network reveals that the lexicon must expand at least as $\approx t^{0.5}$, which corresponds to the observed Heaps' exponent for most languages. This constraint ensures that the exponent of Zipf's Law remains bounded above by $\alpha = 1 / \beta \approx 2$, thereby preserving the ability of language to represent an open-ended conceptual space with a finite vocabulary.

\section{Language network properties}
\label{sec:network}

    \begin{figure*}[t]
    \centering
    \includegraphics[width = 1\linewidth]{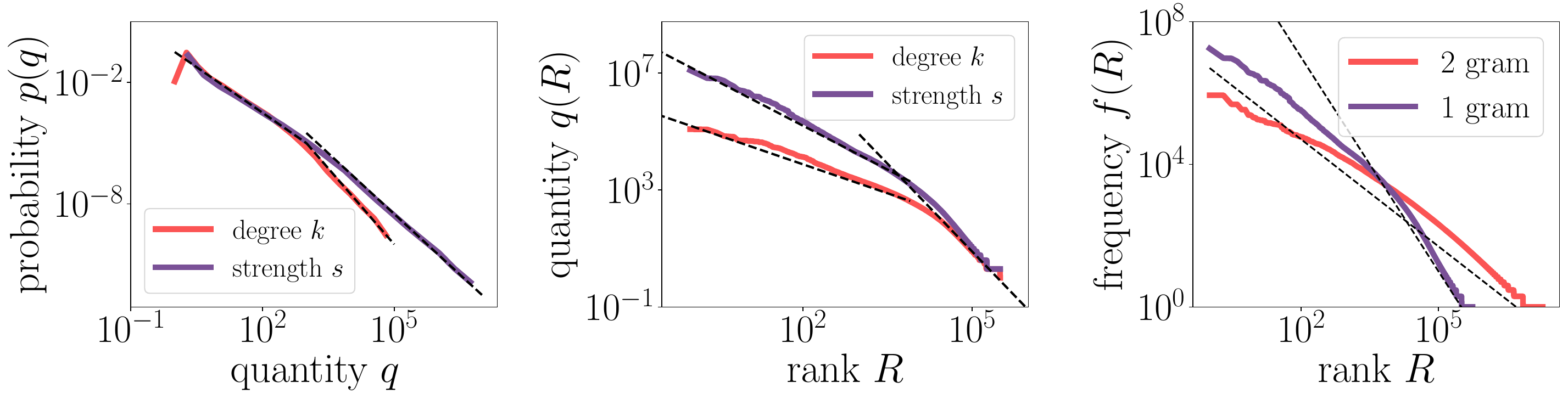}
    \caption{\textbf{Network properties of language.} 
    \textbf{Left:} Degree and strength probability distributions of words in the Word Co-occurrence Network. For small values of degree $k$ and strength $s$ (periphery), both distributions $p(k)$ and $p(s)$ follow a power-law with an exponent of $1.5$ (dashed red line), characterizing a scale-free network with an anomalous exponent. For large values of $k$ and $s$, the exponents shift to $2.5$ and $2$, respectively (see Appendix~\ref{app:C} for details). In the central part of the kernel lexicon, which corresponds to the end of the tail, the degree exponent is close to $\gamma \approx 3$~\cite{cancho2001small}. This value is characteristic of a Barabási-Albert network, indicating high navigability. \textbf{Center:} Rank-size distributions of strength $s(R)$ and degree $k(R)$. Since the strength is equal to word frequency, $s(R)$ follows the same exponent as the frequency-rank distribution $f(R)$, exhibiting a slope of $-1$ in the kernel and $-2$ in the periphery. The degree distribution follows $k(R) \sim R^{-2}$ in the periphery, as $s \sim k$, while in the kernel, $k(R) \sim R^{-2/3}$, reflecting the empirical relationship between degree and strength $k \sim s^{2/3}$~\cite{cancho2001small}. 
    \textbf{Right:} Frequency-rank distributions for 1-grams (single words) and 2-grams (word pairs). The 1-gram distribution transitions to an exponent of $2$ in the tail, corresponding to the periphery of the network. In contrast, the 2-gram distribution maintains an exponent close to $1$ over multiple orders of magnitude of the vocabulary size. This pattern reflects a second-order Heaps' Law, $D_2(t) \approx t$, meaning that the number of new word pairs grows approximately linearly with corpus size. This structural property is intrinsically linked to the lower bound of $2$ observed in the Zipfian distribution of 1-grams. Plots are generated using data from the Project Gutenberg database~\cite{gutenberg}. A summary of the scaling exponents is given in Table~\ref{tab:exponents}. Analyses for other languages are provided in Appendix~\ref{app:D}.}
    \label{fig:fig2}
    \end{figure*}

    Words are organized and co-occur in sentences, revealing their syntactic relationships~\cite{cancho2001small,hanel2020role}. These interactions are essential for creating a structured lexicon, which is needed for meaningful communication. A common way to capture these connections is by analyzing the co-occurrence patterns of words in natural language, which can be represented as a network. In a Word Co-occurrence Network~\cite{kapustin2007vertex, ferrer2007correlations, markosova2006language} (WCN), or Word Adjacency Network, words are nodes, and links represent significant co-occurrences. Typically, a link is placed between every pair of words that are adjacent in the text, which ensures the capture of any possible syntactic relationships~\cite{kapustin2007vertex}. The network is undirected and weighted as any connection is counted with its frequency.

    Like many complex networks~\cite{adamic2000power, wuchty2001scale}, the WCN is largely scale-free, meaning that the degree distribution follows a power law~\cite{cancho2001small, lynn2022emergent, iguchi2005exactly}. A key feature of this network is that the nodes can be divided into two regions, each displaying different scaling exponents~\cite{dorogovtsev2001language}. The ``kernel'' of the network, composed of the approximately 8000 most frequent words for many languages, follows $p(k) \sim k^{-\gamma}$ with $\gamma$ close to $3$, a value typical of Barabasi-Albert networks~\cite{barabasi1999emergence, blumm2012dynamics}. This well-known structure offers great navigability, with a high clustering coefficient and a short average path length between nodes. Every word in the kernel can be reached from any other with fewer than three steps~\cite{barabasi2003scale}. The preferential attachment mechanism ensures that frequent words are more accessible, a phenomenon known as the ``frequency effect''~\cite{akmajian2017linguistics}. This structure optimizes the retrieval of common words and is closely linked to the communication's efficiency within the core lexicon.
    
    In contrast, most less frequent words form a scale-free network with a degree exponent $\gamma \approx 1.5$. Networks with $\gamma < 2$ exhibit key differences~\cite{albert2002statistical, bianconi2001bose, barabasi2001deterministic} from Barabasi-Albert networks. When $\gamma < 2$, the average degree $\langle k \rangle$ diverges with the number of nodes $D$, meaning that the number of links, $D \langle k \rangle/2$, grows faster than the system size. This suggests that forming a link is inexpensive compared to introducing a new node~\cite{seyed2006scale}. A real-world example of such structure is software package networks, where producing a new package is more expensive than using an existing one~\cite{newman2003mixing}. Other examples are peer-to-peer networks like Gnutella~\cite{jovanovic2001modeling}, where each node is a computer, while links are inexpensive logical connections between them. Notably, in both cases, the scaling exponent is close to $\gamma = 1.5$~\cite{seyed2006scale}. Details on scale-free networks and the anomalous regime are reported in Appendix~\ref{app:B}.

    The divergence of the average degree indicates that, at some point in the evolution of language, more connections between words are formed than necessary for optimal communication. As discussed in the previous section, cognitive human limits cause the growth of vocabulary to slow down, necessitating the creation of associations of words to represent complex concepts. To maintain an open-ended conceptual space with a finite vocabulary, it becomes essential to explore more word combinations. Thus, after a highly navigable core lexicon is established, additional word connections are explored, increasing the average degree and reducing overall efficiency. The scaling laws of the Word co-occurrence Network are summarized in Fig.~\ref{fig:fig2}.
    
    Based on these observations, we propose the following framework for the evolution of natural language. Lexicon production and syntactic structures are optimized according to the Principle of Least Effort from information theory. This process organizes the most frequent words into an optimal code, with syntactic connections forming via preferential attachment. This produces a frequency-rank distribution with $\alpha \approx 1$ (Zipf's Law) and a degree distribution with $\gamma \approx 3$, typical of the Barabasi-Albert structure. However, our finite cognitive capabilities prevent the storage of an infinite amount of information in such an efficient way. Once the storage limit is reached, optimization decreases. From a network perspective, we observe that outside the kernel, the network favors the creation of new syntactic connections (edges) over the introduction of new words (nodes). This inefficient recombination process generates the anomalous scale-free region that comprises less frequent words.

    However, vocabulary growth cannot come to a complete halt; it must continue expanding at a sufficient rate to provide enough words for the formation of meaningful combinations. We can analyze this limit by studying the scaling of the number of links in the Word Co-occurrence Network. This grows approximately linearly with the corpus size, following $D_2(t) \approx t$ over several orders of magnitude (Fig.~\ref{fig:fig2}). In any network, the number of nodes must scale at least as the square root of the number of links~\cite{di2025dynamics}. This implies that vocabulary growth must be greater than or equal to $D_1(t) \approx t^{0.5}$. As a consequence, Zipf’s exponent {$\alpha = 1 / \beta$} remains bounded above by $\alpha \approx 2$, ensuring that language can continue to represent an open-ended conceptual space with a finite vocabulary. We provide a formal derivation of this bound on Zipf's exponent in Appendix~\ref{app:limit}.

\section{A generative model for the evolution of language}
\label{sec:model}
    
    In this section, we introduce a generative model that describes the statistical properties of language by integrating the Adjacent Possible concept~\cite{kauffman1996investigations, tria2018zipf, loreto2016dynamics} with communication optimality loss.
    
    The model we propose is a Time-Dependent Urn Model with Triggering~\cite{bellina2024time, tria2014dynamics, polya1930quelques, chen2013generalized, pemantle1990time}, which can also be interpreted as a network generative process~\cite{di2025dynamics, iacopini2018network} for vocabulary evolution. In this representation, words correspond to nodes, while links capture syntactic relationships between them. At $t=0$, the network consists of an initial set of $n_0$ nodes. At each time step, a node is selected randomly, and a link is created between consecutive extractions in the sequence.

    Following the ideas of urn models, we can conceptualize this network as composed of two spaces. The \textit{actual space} (or \textit{actual network}) includes all nodes that have been selected at least once, corresponding to words that are already part of the vocabulary. Its size is denoted as $D(t)$. The \textit{adjacent possible} consists of nodes that exist in the system but have not yet been explored~\cite{kauffman1996investigations, loreto2016dynamics}, representing words that have not yet been used but are accessible in one step. At each time step, a word is either recalled from the actual space or introduced as a novelty from the adjacent possible. The latter event corresponds to introducing a new word into the vocabulary. The probability of recalling from the actual network, $p_{\text{act}}$, is proportional to the total weight of the links in the network, while the probability of introducing a novel word is $p_{\text{adj}} = 1 - p_{\text{act}}$. The explicit expressions are reported in Appendix~\ref{app:C}.

    In the actual network, a word $i$ is selected with a probability directly proportional to its strength:
    \begin{equation}
        \text{prob}_i(t) = \frac{s_i(t)}{\sum_{j=1}^{D} s_j(t)}
        \label{eq:prob_actual}
    \end{equation}
    Each time a word is extracted, it forms a link with the previously selected ones, increasing its strength $s_i$ by one. If the connection already exists, its weight is incremented by one as well. This mechanism generates a network which is directly analogous to Word Co-occurrence Networks constructed from textual data. The process described in Eq.~\ref{eq:prob_actual} shares similarities with the Barabási-Albert preferential attachment model~\cite{barabasi1999emergence}, with selection probability depending on strength rather than degree. This leads to the creation of a highly optimized kernel lexicon, ensuring efficient navigation and communication.
    
    The probability of selecting a word from the adjacent possible, which corresponds to introducing a novelty, is proportional to the size of this space. To allow for network growth, each newly discovered word expands the adjacent possible by one unit, following the triggering mechanism of Urn Models with Triggering~\cite{tria2014dynamics}. These newly added nodes remain unconnected until they are explored for the first time. Consequently, the size of the adjacent possible coincides with the number of novel words introduced, which is simply the vocabulary size $D(t)$. As a result, the probability of introducing a new word evolves as:
    \begin{equation}
        \frac{dD(t)}{dt} = \frac{n_0 + D(t)}{n_0 + D(t) + t} \approx \frac{D(t)}{D(t) + t}
    \end{equation}
    where the normalization accounts for the size of the adjacent possible $D(t)$, the initial number of nodes $n_0$, and the total weight of the links in the actual network $t$. This mechanism ensures the formation of an optimal structure, leading to a frequency-rank distribution that follows Zipf's Law, $f(R) \sim R^{-1}$ (see Appendix~\ref{app:C}). Moreover, in the actual network, the strength of a node is directly related to its frequency, being exactly twice as large.
        
    \begin{figure}[t]
    \centering
    \includegraphics[width=\linewidth, trim = 20 160 20 160, clip]{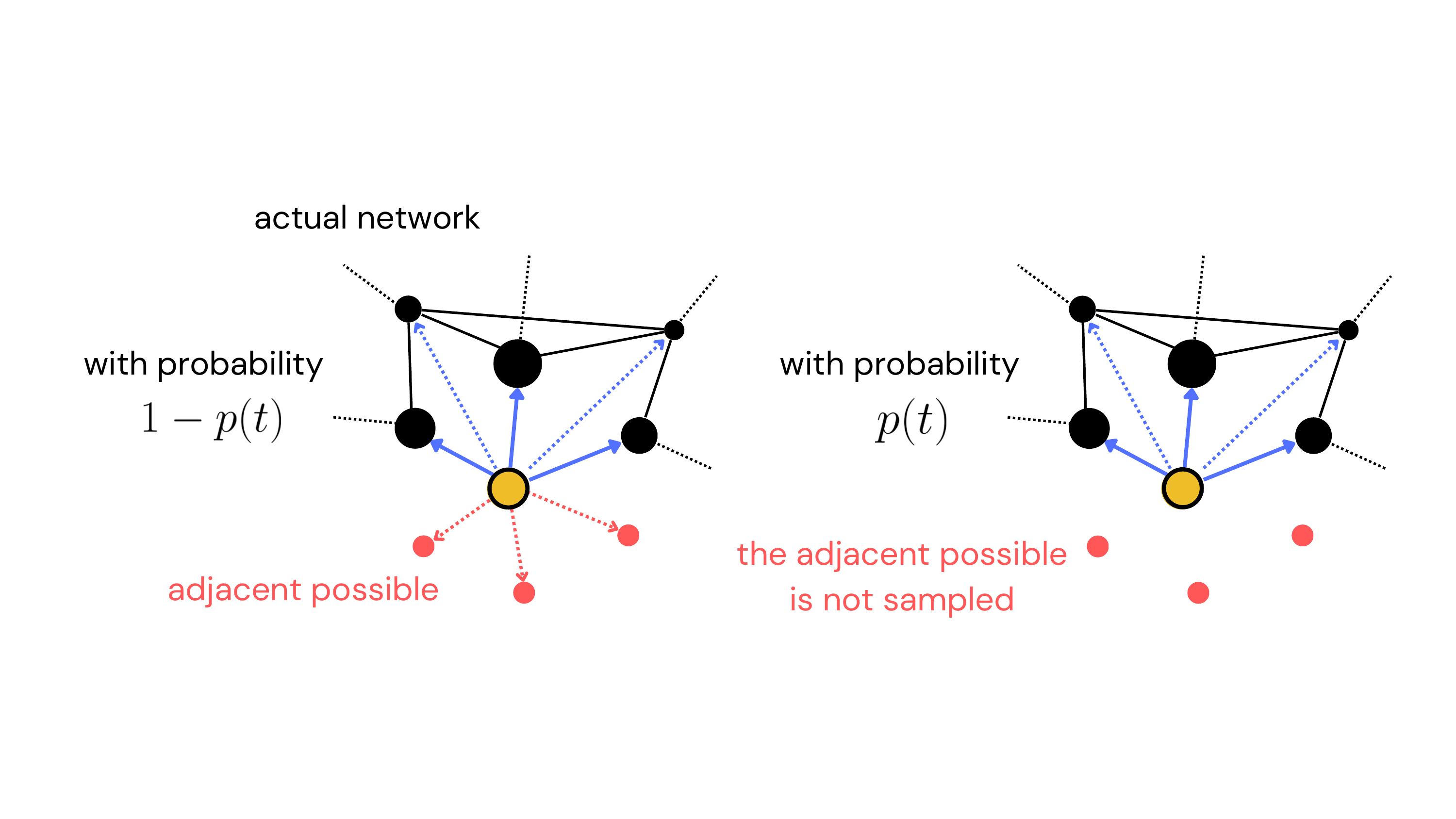}
    \caption{\textbf{Schematic representation of the network generative model.} The model dynamics consist of an edge-reinforced random walk~\cite{sabot2011edge} on a network, combined with the triggering mechanism typical of urn models with triggering~\cite{tria2014dynamics}. In this model, nodes represent words while edges are their syntactic connections. The size of each node represents its strength. The network comprises the actual space, made up of all nodes that have already been visited, and the adjacent possible space of nodes that have never been visited. Every time a new node is discovered, the adjacent possible space is enlarged by one element (triggering mechanism). Starting from the yellow node, we select the following node of the walk with a probability proportional to the strength of this node. With probability $p(t)$, we neglect the presence of the adjacent possible, i.e., the possibility of discovering a new node. This time-dependent mechanism, typical of non-stationary urn models~\cite{bellina2024time}, represents the diminishing efficiency in language production.}
    \label{fig:fig4}
    \end{figure}

    The final ingredient we introduce is a mechanism that reduces the probability of adding new words in favor of creating associations between existing ones. As observed earlier, cognitive limits slow down vocabulary growth, favoring the formation of more syntactic associations. We model these cognitive limits through a term $p(t)$, which regulates the probability that an association is formed between words already present in the vocabulary, without considering the adjacent possible (Fig.~\ref{fig:fig4}). Consequently, as vocabulary grows, the probability of introducing a new word from the adjacent possible gradually decreases as:
    \begin{equation}
        \dfrac{d D(t)}{dt} \approx (1 - p(t)) \dfrac{D(t)}{D(t) + t}.
    \label{eq:model}
    \end{equation}
    We choose the specific form of $p(t)$ to depend on the number of pairwise combinations that can be formed from the vocabulary, which {at most} scales as $D(t)^2$. A sigmoid-like function works well for this purpose:
    \begin{equation}
        p(t) = \dfrac{1}{2} \dfrac{D(t)^2}{D_0^2 + D(t)^2}.
    \end{equation}
    However, the exact form of $p(t)$ is not crucial for the model's behavior. The important property is that it must start at $p(0) = 0$ and saturate at $p(t) \to 1/2$ for large $t$. The factor $1/2$ is the upper bound for this probability, above which no new associations would be available to explore for large times. The function $p(t)$ ensures that vocabulary growth remains sufficiently high (above $D(t) \sim t^{0.5}$) to allow the continuous creation of new word associations, and its asymptotic value is determined by the constraints discussed in the previous sections.
    
    The resulting network has a core organized as a scale-free network, with a degree exponent $\gamma$ between $2$ and $3$. The model reproduces the observed scaling of $p(k) \sim k^{-1.5}$ for less common words. Equation~(\ref{eq:model}) predicts deviations from Zipf's Law outside the kernel, with $\alpha \approx 2$, which is the lower bound needed to maintain new syntactic combinations, as seen in real language. Finally, the model reproduces Heaps' Law with $\beta = 0.5$. Additional details on the scaling predicted by the model and results of numerical simulations are given in Appendix~\ref{app:C}.

    The quantity $D_0$, which is the physical parameter of the model, can be related to the cognitive capabilities of humans. Specifically, $D_0$ represents the size of the kernel lexicon, which is the maximum number of words that can be optimized for efficient communication. This value can be interpreted as a measure of the brain's storage capacity, which allows for the optimization of only a finite number of words. Once the lexicon exceeds this limit, inefficiencies begin to emerge, reducing the overall navigability of the system and slowing down vocabulary growth.
    
    We can measure $D_0$ by fitting the model to a real corpus of text, obtaining $D_0 \approx 8000$, which aligns with previous estimates of the core lexicon size for the English language~\cite{cancho2001small, gerlach2013stochastic}. To test the hypothesis that $D_0$ is related to cognitive capabilities, we compare texts produced by native English speakers (L1 corpus) with those written by non-native speakers learning the language (L2 corpus~\cite{lozano2020designing}). We show the results in Fig.~\ref{fig:fig3} in the Appendix~\ref{app:C}. The L2 corpus exhibits a less efficient use of the language, reflected in a smaller size of the kernel lexicon. Having a smaller vocabulary, learners rely heavily on combinations of words to express complex ideas, producing longer texts with fixed information content. The value measured in the L2 corpus is $D_0 \approx 200$, highlighting the reduced optimization capabilities of learners compared to native speakers. It is important to stress that this interpretation is meaningful only for intra-language comparisons. When comparing different languages, differences in $D_0$ must be interpreted with caution. For instance, German displays a significantly larger kernel lexicon ($D_0 \approx 20000$) compared to other languages~\cite{gerlach2013stochastic}, but this does not indicate greater efficiency. It rather reflects the language’s strong tendency to form compound words~\cite{fleischer2012word}, which artificially inflates the vocabulary size, as discussed in Appendix~\ref{app:C3}.

\section{Discussion}

    The statistical properties of language have always attracted great interest in the literature, revealing remarkable characteristics that serve as a footprint of the complex processes governing its evolution. Since the early studies by George K. Zipf~\cite{zipf2013psycho}, the driving force in the evolution of language was recognized in the attempt to communicate information as efficiently as possible. The formalization of the Principle of Least Effort~\cite{cancho2003least, i2005decoding, agouzal2024relationship, cubero2019statistical} has demonstrated that the ubiquitous presence of power laws in language statistics emerges as a manifestation of statistical criticalities in the communication process.

    However, language optimization is inherently constrained by the cognitive limitations of the human brain. This constraint manifests in the transition of Zipf's exponent from $1$ to $2$, suggesting that cognitive boundaries prevent full vocabulary compression and optimization. This phenomenon becomes particularly evident in Word Co-occurrence Network analysis, which highlights structural constraints that establish $2$ as a upper bound for the exponent.
    
    This research integrated three distinct domains of complex systems science to address these phenomena. First, we apply principles from information theory to examine the relationship between word frequency-rank distributions in texts and communication efficiency. Second, we leverage complex network theory to establish structural bounds for the frequency-rank exponent of rare words, offering an interpretation of its asymptotic convergence to -2. Finally, we introduced a model based on P\'olya's urn and the adjacent possible, which successfully reproduces the observed frequency-rank distributions in texts.
    
    The limited capacity of the human brain inhibits the growth of vocabulary, thus reducing expressive capacity and necessitating the use of word combinations, which in turn leads to an increase in text length. In principle, if artificial systems, such as large language models (LLMs), possess greater storage capacity than humans, they could accommodate a larger vocabulary, allowing more concepts to be associated with single words. Our findings suggest a practical approach to improving communication efficiency: identifying groups and combinations of words associated with single concepts and replacing them with a single codeword.  This could help improve linguistic capabilities by significantly reducing the number of tokens used in text generation. This phenomenon is observed in the German language, which displays a smaller Zipf's exponent ($1.5$ instead of $2$) because it replaces complex combinations of words with single expressions. However, the limitation of German is that it often concatenates words without creating genuinely new and shorter terms, which reduces its advantage in terms of text length when considering the number of letters. Further research should investigate the size of the kernel lexicon in corpora produced by LLMs to determine whether the level of efficiency in their language is higher, lower, or comparable to that of humans. 
    
    The idea that variations in Zipf's exponent can differentiate writers with different cognitive or linguistic capabilities dates back to the first studies in linguistics~\cite{zipf2013psycho} and has been extensively investigated~\cite{ferrer2005variation}. Notably, individuals affected by schizophrenia tend to exhibit larger Zipf's exponents. Our framework provides a direct explanation for these observations: these individuals show a disordered and less efficient use of language, which manifests as a smaller kernel lexicon, revealing the exponent $2$ much earlier than in typical speakers. In our framework, linguistic disorders and differences among speakers (such as the case of English learners reported in Section~\ref{sec:model}) can be detected by measuring the size of the kernel lexicon $D_0$. This approach allows for a direct comparison and differentiation of corpora produced by different speakers. However, cross-language comparisons must be treated with caution, as the size of the kernel lexicon can be influenced by structural properties of the language. In the case of German, for instance, the frequent use of compound words leads to a larger kernel lexicon without a corresponding gain in communicative efficiency.

    A further and more sophisticated analysis would require studying language at the semantic level. In our analysis, we considered only the lexical level, treating each word as a sequence of symbols without connecting them to specific concepts. Analyzing the semantic content of words through embedding procedures, for example, would provide direct access to a representation of the concept space, allowing us to understand how words, and more importantly, combinations of words, are mapped into meanings and ideas. In conclusion, this work opens the way for a deeper understanding of the mechanisms and principles governing the evolution of language and provides a foundation for designing more effective communication techniques.

\vspace{6pt} 


\paragraph{Author contributions:}
Idea, V.D.P.S.;
Conceptualization, V.D.P.S. and A.B.; 
methodology, A.B. and V.D.P.S.; 
software, A.B.; 
validation, A.B.; 
formal analysis, A.B.; 
investigation, A.B.; 
resources, A.B. and V.D.P.S.; 
data curation, A.B.; 
writing---original draft preparation, A.B.; 
writing---review and editing, A.B. and V.D.P.S.; 
visualization, A.B.; 
supervision, V.D.P.S.; 
project administration, V.D.P.S.; 
funding acquisition, V.D.P.S.;
All authors have read and agreed to the published version of the manuscript.

\paragraph{Funding:}
This research was financed by funds from the Austrian Federal Ministry for Climate Action, Environment, Mobility, Innovation \& Technology (BMK) as part of the CSH Postdoc Programme (GZ 2023-0.841.266). 

\paragraph{Acknowledgments:}
We acknowledge valuable discussions with M.~Marsili, S.~Thurner, E.~G.~Altmann, and G.~De~Marzo.

\paragraph{Conflicts of interest:}
The authors declare no conflict of interest.

\newpage

\appendix

\section{Optimal information compression and statistical criticalities}
\label{app:A}

    \subsection{Maximization of the relevance at constant resolution}

        In this section, we explain the framework of optimal information compression and statistical criticalities, based on~\cite{cubero2019statistical, haimovici2015criticality, marsili2013sampling}. The framework is adapted here to describe linguistic phenomena, focusing on the trade-off between relevance and resolution in human language.
        
        Consider a generative process that produces a sample of $N$ objects, denoted as $\hat{x} = {x_1, \dots, x_N}$, of any type (mathematical or otherwise). We define a representation as a set $\hat{s} = {s_1, \dots, s_N}$, drawn from some alphabet $\mathcal{S}$, such that each $x_i$ is represented by a symbol $s_i$. The number of available symbols, $D=|\mathcal{S}|$, can be smaller than $N$. We refer to $D$ as the vocabulary, or dictionary. For the same data $\hat{x}$, we can choose different representations $\hat{s}$ with varying levels of detail.
        
        In the context of language, the generative process corresponds to the act of communication between a speaker and a hearer, where the speaker aims to convey specific concepts or meanings. The elements $\hat{x} = {x_1, x_2, \dots }$ are therefore the (virtually infinite) concepts or ideas that the speaker wants to communicate to the hearer. The representation we choose to describe this data is what we refer to as the vocabulary: the set of words used to describe and convey these concepts. Each concept $x_i$ is mapped to a finite vocabulary $\hat{w} = {w_1, w_2, \dots, w_D}$ of words, where $D$ represents the total number of unique words available. Since the set of possible meanings is infinite, conveying all possible concepts with a finite vocabulary requires forming combinations of words $w_i$~\cite{goldberg1995constructions, croft2001radical}. In this way, the language acts as a function $W(x)$, where $x$ represents a given concept, and $W(x)$ is the set of words used to describe it. The frequency of occurrence of a word $w$ in a corpus is denoted by $f_w$, and its rank by $R_w$.
        
        The efficiency of communication depends on maximizing \textit{relevance} (the informativeness of the words) while minimizing \textit{resolution} (the cognitive effort required to distinguish and use different words)~\cite{marsili2022quantifying, cubero2018minimum}. Resolution is defined as the information content of the sample when represented by $\hat{s}$, measuring the number of bits needed to describe one of the outcomes. In language context, the resolution represents the cognitive effort required to distinguish between different words in the vocabulary $\hat{w}$. It can be mathematically expressed as: 
        \begin{equation} 
            H[w] = - \sum_{w=1}^{D} \dfrac{f_w}{N} \log \dfrac{f_w}{N}, 
            \label{eq:resolution}
        \end{equation}
        where $f_w$ is the frequency of word $w$, i.e., the number of occurrences of $w$ in the text of length $N$.
        The resolution is the Shannon entropy of the text.
        
        Relevance, on the other hand, represents the maximal amount of information (in bits) that can be extracted from the representation $\hat{w}$ about the underlying generative process. It quantifies the extent to which the hearer can extract information from the message about the underlying meanings being communicated. Relevance is defined as: 
        \begin{equation} 
            H[f] = -\sum_{f=1}^N \dfrac{f m_f}{N} \log \dfrac{f m_f}{N}, 
            \label{eq:Hf}
        \end{equation} 
        where $m_f = \sum_{w=1}^{D} \delta_{f,f_w}$ represents the number of words $w$ that appear with frequency $f$. 
        The relevance is therefore the Shannon entropy of the frequency probability distribution of words.
        Rewriting the resolution in terms of frequency, we have: 
        \begin{equation} 
            H[w] = - \sum_{f=1}^N \dfrac{f m_f}{N} \log \dfrac{f}{N}. 
            \label{eq:Hw}
        \end{equation} 
        The difference between resolution and relevance can be expressed as: 
        \begin{equation} 
            H[w] - H[f] = H[w|f] = \sum_{f=1}^N \dfrac{f m_f}{N} \log m_f, 
        \end{equation} 
        which quantifies the level of noise in the sample—those bits of information that cannot be used to describe the underlying generative process accurately. This noise reflects inefficiencies or ambiguities in communication, which reduce the clarity of the information being conveyed.
        
        In the context of language, $H[w]$ represents the speaker's effort, as it quantifies the number of bits needed to encode the vocabulary effectively. The speaker aims to compress this information as much as possible. On the other hand, $H[f]$ quantifies the amount of information that the hearer can extract from the message. Thus, the trade-off between resolution and relevance is a balance between achieving maximal compression without compromising the hearer's ability to understand the conveyed concepts.
        
        To illustrate this trade-off, consider two extreme cases. In the first case, we use a representation with the finest level of description, where each concept is assigned a distinct word ($w(x_i) \neq w(x_j)$ for all $i \neq j$). In this situation, the resolution is maximized, $H[w] = \log N$, as the vocabulary is infinitely large with a unique word for each concept. However, the relevance $H[f]$ is zero since $m_1 = N$ and $m_f = 0$ for any $f > 1$. This implies that if every word appears only once, the hearer cannot statistically discern between them, resulting in no informative gain from the message~\cite{cubero2019statistical}. The other extreme case corresponds to the lowest level of description, where all concepts are represented by the same word ($w(x_i) = w_0$ for all $i$). Here, the cognitive effort of the speaker in terms of resolution is $H[w] = 0$. Since $m_N = 1$ and $m_f = 0$ for all $f \neq N$, the relevance $H[f]$ is also zero, because the hearer gains no useful information from the message if the same word represents every concept. 
        
        Intermediate values of $H[w]$ correspond to varying levels of detail in the representation. The goal is to find a balance where the representation is compressed enough to reduce the cognitive effort of the speaker while retaining as much relevance as possible for the hearer. We can formulate the communication process as an optimization problem where the goal is to maximize $H[f]$ while keeping $H[w]$ constant, thus maximizing useful information at a fixed level of compression. Mathematically, this is represented by the following Lagrangian problem:
        \begin{equation}
            \Gamma = H[f] + \mu (H[w] - H_w) + \lambda \left(\sum_w f_w - N\right).
            \label{eq:lagrangian}
        \end{equation}
        This problem is optimized for power laws of the form:
        \begin{equation}
            m_f \sim f^{-1-\mu},
        \end{equation}
        where the Zipf's exponent is given by $1/\mu$. Zipf's Law ($\mu = 1$) is obtained when the relevance is maximally preserved while compressing information without loss. These representations are called \textit{maximally informative representations} and enable an optimal exchange of information between the speaker, who represents the underlying process, and the hearer, who decodes it~\cite{cubero2019statistical}. 
        
        The Lagrange multiplier $\mu$ represents the slope of the trade-off between resolution and relevance. In other words, in the compression process, each bit sacrificed in resolution results in $\mu$ bits of relevance. Compression is beneficial and lossless as long as $\mu > 1$, since each bit sacrificed in resolution is transformed into more bits of relevance, implying no loss of information. Conversely, when $\mu < 1$, compressing by one bit in resolution does not fully convert into relevance, resulting in a loss of information. Thus, $\mu=1$ represents the point at which the representation is maximally compressed without loss, and it is also the point where $H[f] + \mu H[w]$ is maximal. 
        
        Frequent words are associated with high relevance and low cognitive cost, forming the ``kernel'' of the language~\cite{cancho2001small}. This kernel vocabulary is optimized for efficient communication, minimizing cognitive effort for both the speaker and the listener, which is why we observe an exponent of $1$ in the frequency-rank distribution. However, as we move toward less frequent words, cognitive constraints become more pronounced. The brain cannot accommodate an indefinitely growing quantity of information, which is measured by $H[w]$. With a frequency rank distribution with an exponent of $1$, this quantity diverges, which is not feasible due to cognitive limits~\cite{miller1963finitary, lu2010zipf, newman2005power}. More precisely, from Eq.~(\ref{eq:resolution}), the resolution can be rewritten in terms of the ranks as $H(w) = -\sum_{R_w=1}^D \dfrac{f_{R_w}}{N} \log \dfrac{f_{R_w}}{N}$. This quantity diverges as $D$ increases when $f_R \sim R^{-1}$ and converges for $f_R \sim R^{-\alpha}$ with any $\alpha > 1$. Thus, a bending of the exponent to values larger than $1$ is expected to ensure that the amount of information remains finite.
        
        To maintain an open-ended conceptual space with a finite vocabulary, the system sacrifices resolution, leading to the use of combinations of common words to express nuanced or novel ideas. This is reflected in the frequency distribution steepening to an exponent of $2$, which represents lossy compression where the system loses resolution to accommodate emerging concepts. Thus, concepts are not directly mapped into words, but more in general into n-grams (combinations of words). Given the optimization required in the communication process, we expect that the n-grams are distributed according to Zipf's Law, indicating that concepts are being represented efficiently through combinations of words, as explained in Fig.~\ref{fig:fig1}.

\subsection{Interplay between resolution and relevance}
\noindent

\begin{figure}
    \centering
    \includegraphics[width=0.48\linewidth]{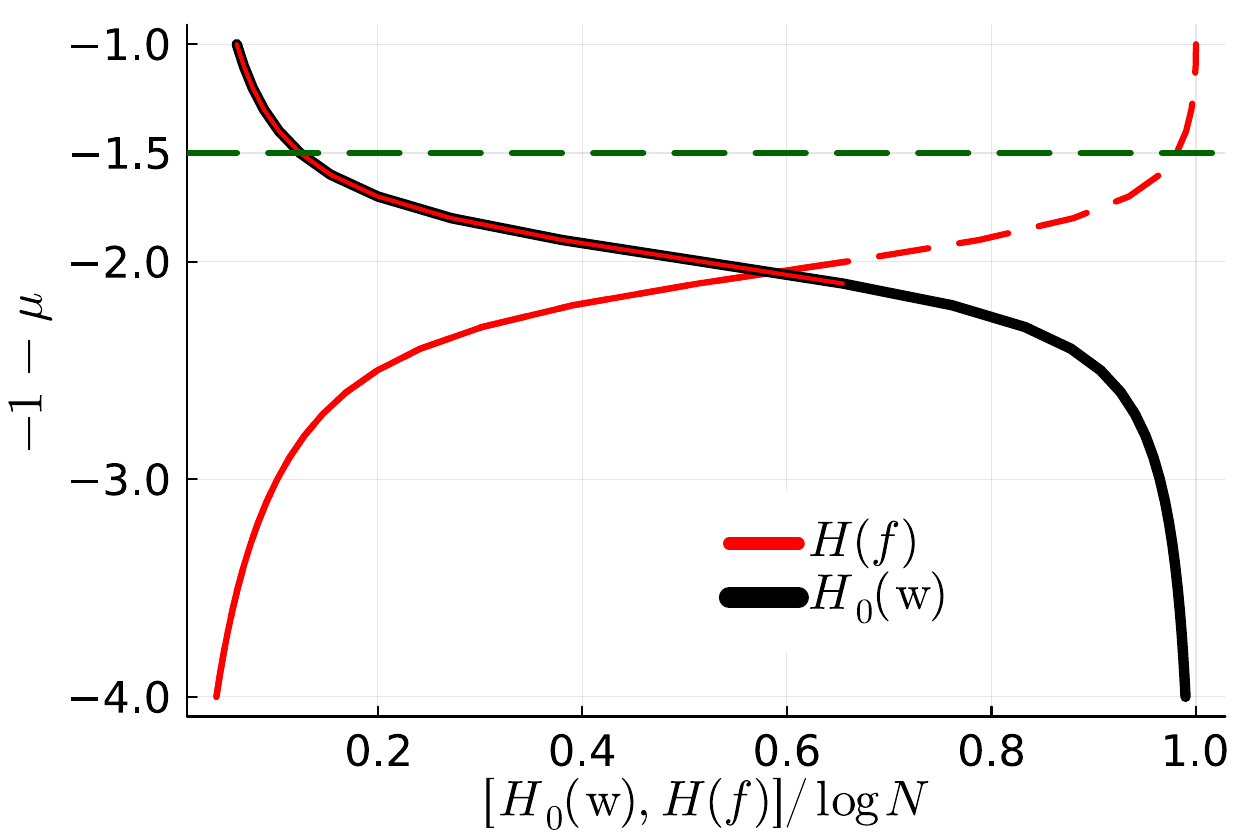}%
    \includegraphics[width=0.48\linewidth]{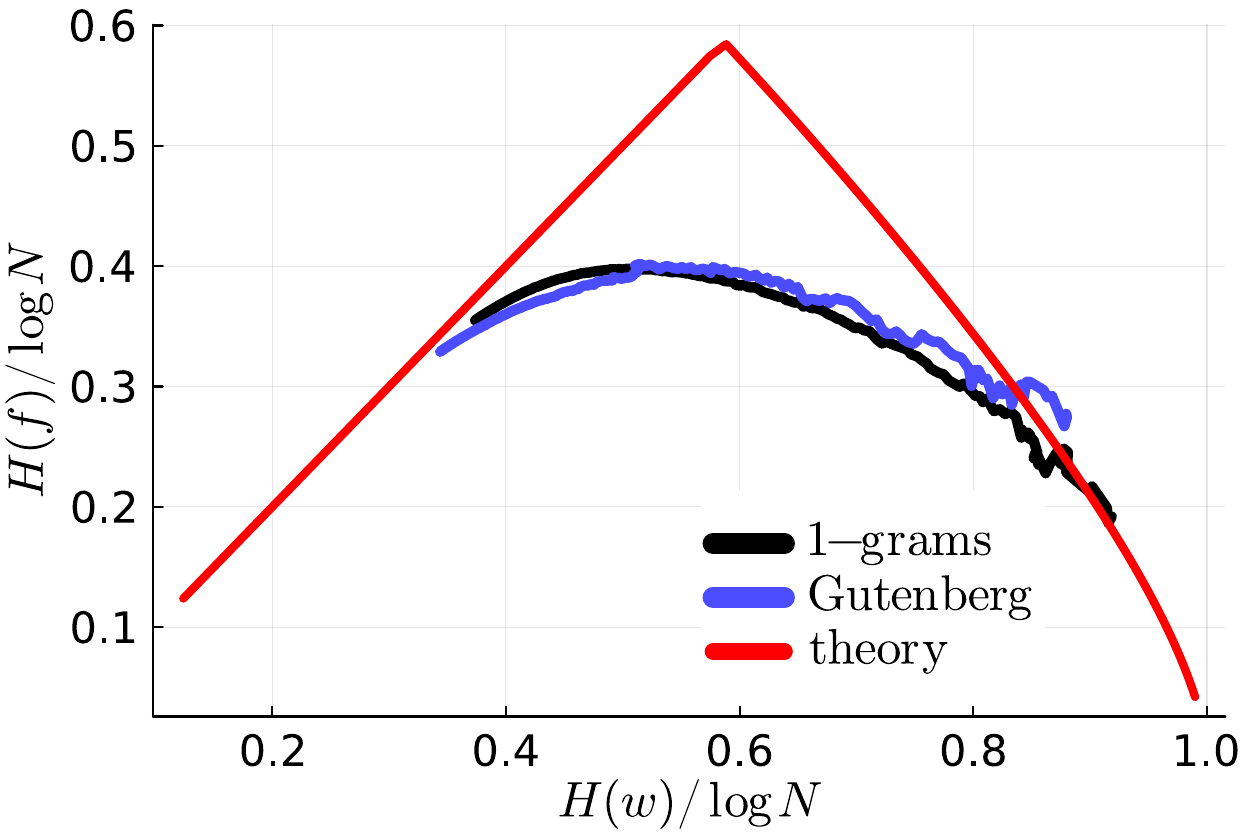}%
    \caption{\textbf{Interplay between resolution and relevance.}
    \textbf{Left:} 
        The exponent of the power-law frequency distribution (vertical axis) shows the relationship between normalized resolution $H_0(w)$ (black line) and relevance $H(f)$ (red lines) (horizontal axis). 
        The curves are calculated using Eqs.~(\ref{eq:theoryentropies}) that consider $m_f\propto f^{-1-\mu}$, normalization constraint $\sum_f fm_f=N$, and $N=10^7$.
        The critical value $\mu\sim 1$ distinguishes between lossy and lossless compression regions. 
        When $\mu<1$, the number of elements $N$ becomes too small to accurately represent the frequency distribution (dashed red line). 
        The case $H(f)=\log N$, for example, would require $N$ different frequencies with $N$ elements, which is impossible (undersampling).
        In this specific case, $\mu=0$ and the system exhibits many identical elements with higher frequencies, leading to very low resolution.
        The dashed green line corresponds to $\mu=0.5$ and thus to a Zipf exponent of $-2$.
    \textbf{Right:}
        Theoretical dependence between resolution and relevance derived from the left panel (red line) is compared with data from a stream of $10^9$ words sampled from the 1-grams word frequency (black line) and with the stream of Gutenberg containing $3.3\times 10^8$ words (blue line). The black curve peaks at approximately $3.5\times 10^6$ total words, corresponding to $1.4\times 10^5$ unique words.
        The green line peaks at approximately $1.7\times 10^5$ with $1.1\times10^4$ unique words.
    }
    \label{fig:interplay}
\end{figure}

To explore the interplay between resolution (Eq.\ref{eq:Hw}) and relevance (Eq.\ref{eq:Hf}), we perform a numerical analysis focusing on power-law frequency distributions of the form $m_f \propto f^{-1-\mu}$. These distributions are normalized to satisfy $\sum_f m_f f = N$, where $N$ denotes the total number of elements in the system, e.g., the total number of words in a text.
Such a power-law distribution emerges from the optimization of relevance at a fixed resolution. Building on this insight, we evaluate both $H_0(w)$ and $H(f)$ as functions of the power-law exponent $\mu$ using:
\begin{equation}
H_0(w) = -\frac{\sum_{f=1}^N f^{-\mu}\log \frac{f}{N}}{\sum_{f=1}^{N}  f^{-\mu}}
~~~\mbox{and}~~~
H(f) = -\frac{\sum_{f=1}^N f^{-\mu}\log \frac{f^{-\mu}}{\sum_{f=1}^{N}  f^{-\mu}}}{\sum_{f=1}^{N}  f^{-\mu}},
\label{eq:theoryentropies}
\end{equation}
which we obtain from Eqs.~(\ref{eq:Hw}) and (\ref{eq:Hf}) by inserting $m_f = Nf^{-1-\mu}/\sum_f f^{-\mu}$. We consider a system of $N = 10^7$ elements and plot both quantities against $-1-\mu$ in the left panel of Fig.~\ref{fig:interplay}. Fixing the value of $H_0(w)$ corresponds to setting a human cognitive limit at a given resolution, effectively restricting the number of available microscopic states $\Gamma \sim e^{N H_0(w)}$. This constraint could be represented as a vertical line in the plot; its intersection with the black curve would determine the exponent $-1-\mu$, which in turn, moving horizontally to the left to intersect the red curve,  defines the corresponding value of $H(f)$. 

To help interpret the left panel of Fig.~\ref{fig:interplay}, let's consider a sequence of words with a high relative resolution of $0.8$. This value corresponds to a frequency distribution following $m_f \propto f^{-2.2}$ ($\mu = 1.2$) and results in a Zipf exponent of approximately $1 / \mu = 0.8$. Given the high resolution, the system can effectively manage a frequency distribution with an exponent smaller than $1$, allowing for more frequent word usage compared to a standard Zipf law. Conversely, with a lower relative resolution of $0.2$, the frequency distribution approaches $m_f\propto f^{-1.7}$, yielding a Zipf exponent of $1.4$ (exceeding unity). In this regime, the system operates in a lossy compression domain, where relevance is constrained by resolution due to the fundamental requirement that relevance cannot exceed resolution.

The right panel of Fig.~\ref{fig:interplay} explicitly shows the relevance as a function of resolution. The red line represents the theoretical expectation, obtained with $N = 10^7$.
Starting from maximum resolution, $H(w) / \log N = 1$, the system sacrifices bits in resolution to gain relevance. In this region, each bit compressed in resolution translates into $\mu$ bits of relevance. As long as $\mu > 1$, compression remains lossless and beneficial. However, after reaching the optimal point $\mu = 1$, where the representation is maximally compressed while retaining the highest relevance, further compression becomes lossy. The black and blue lines represent empirical data from the Google 1-gram~\cite{michel2011quantitative} and Project Gutenberg~\cite{gutenberg}. While the theoretical and empirical curves do not align perfectly due to simplifications in the model, they exhibit similar qualitative features, such as the presence of a maximum, the rightward-to-leftward trajectory as $N$ increases, and the tendency for resolution and relevance to converge at large $N$. It is worth noting that sampling from the 1-gram distribution effectively reflects collective linguistic knowledge rather than an individual's language pattern.

\begin{figure}
    \centering
    \includegraphics[width=.8\textwidth]{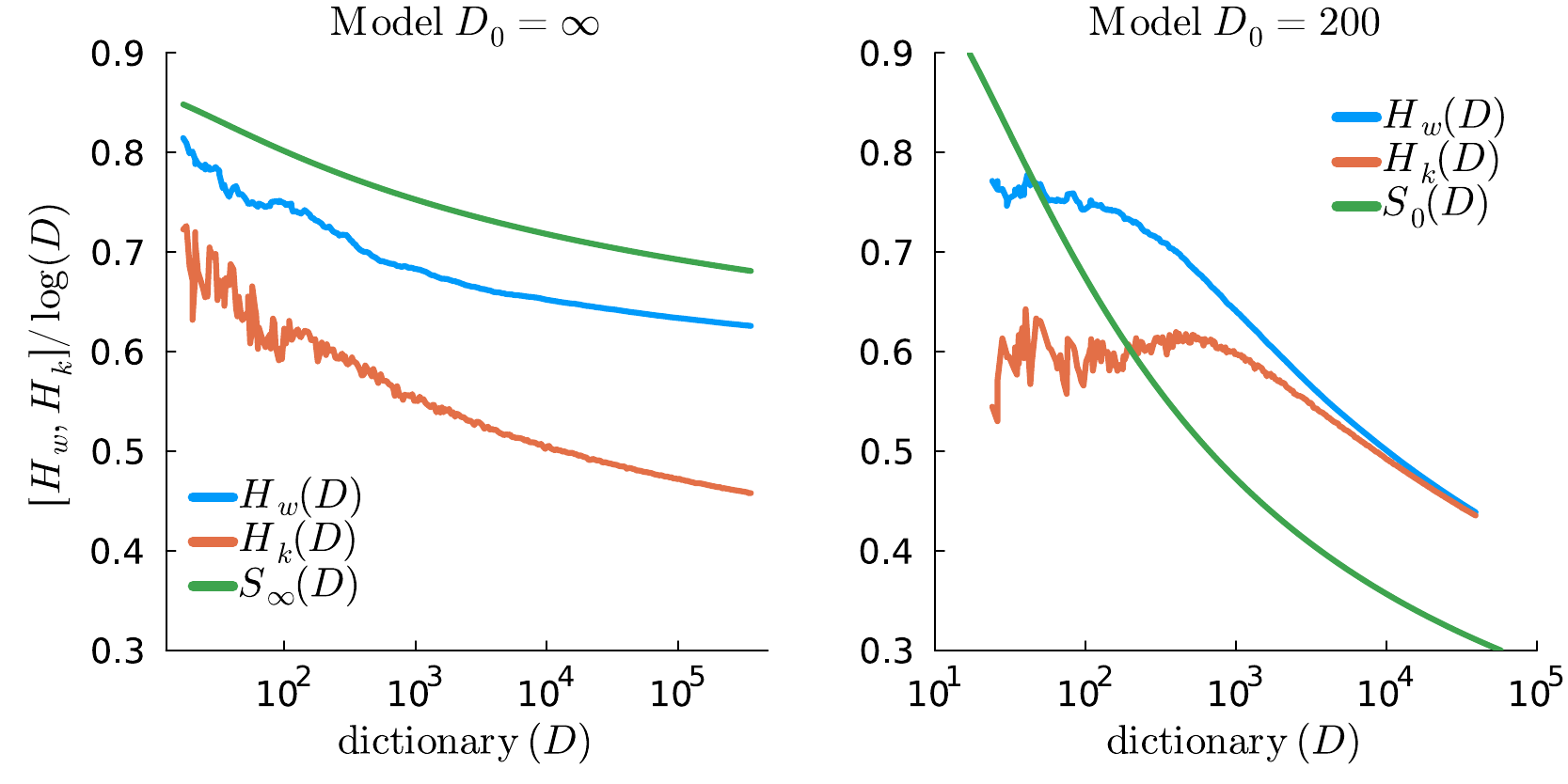}
    \caption{\textbf{Resolution and relevance dynamics in the generative model.}
        The evolution of normalized resolution and relevance is plotted as a function of unique tokens in the generated stream.
        \textbf{Left:}
            Entropy trajectories for the unconstrained model ($D_0=\infty$), with the green line representing the theoretical resolution at $\mu=1$ (Zipf exponent equals 1).
        \textbf{Right:}
            Entropy trajectories under strong cognitive constraints ($D_0=200$), with the green line showing the theoretical resolution at $\mu=0.5$ (Zipf exponent equals 2).
        Both configurations demonstrate the same asymptotic behavior of the resolution and the theory at large $D$.
        }
    \label{fig:model_entropies}
\end{figure}

We further extend our analysis to explore the temporal evolution of resolution and relevance within our generative urn-based model. By tracking these quantities over time during stream generation, we infer their dynamic behavior, as shown in Fig.~\ref{fig:model_entropies}. To establish theoretical benchmarks, we derive resolution curves from Equation~\ref{eq:theoryentropies} using a continuous approximation, considering the addition of unique words at each step. This approach replaces discrete summations $\sum_{r=1}^D f(r)$ with the corresponding integrals $\int_1^D f(r)\,dr$. We refine this approximation by incorporating known asymptotic relations for large $D$: $\sum_1^D \frac{1}{r} \approx \gamma + \log D, \ \sum_1^D \frac{1}{r^2} \approx \frac{\pi^2}{6}$, where $\gamma$ denotes the Euler-Mascheroni constant.
The theoretical estimates of the resolution in the cases $D_0 = \infty$ (unconstrained) and $D_0 = 200$ (strong cognitive constraint) take the following forms:
\[
S_\infty(D) = \frac{1}{2}\frac{\log D}{\gamma + \log D} + \frac{\log(\gamma + \log D)}{\log D},
\]
\[
S_0(D) = \frac{\pi^2}{3} \cdot \left( \frac{1}{\log D} - \frac{1}{D} - \frac{1}{D \log D} \right).
\]

Although the theoretical curves exhibit some deviations from measured values, these discrepancies can be attributed to specific factors in each case. In the unconstrained scenario ($D_0=\infty$), the deviation arises from the gradual emergence of Zipf’s law during the generation process. In the cognitively constrained case (small $D_0$), additional divergence occurs because the theoretical curve assumes $D_0=0$, corresponding to an instantaneously established Zipf exponent of $-2$. Nevertheless, the observed asymptotic convergence in both scenarios validates our theoretical framework for predicting long-term behavior.

    \subsection{The Limits of lossy compression}
        \label{app:limit}
        When the exponent of the frequency distribution shifts to $2$, the system transitions into a regime of lossy compression. This marks a point where the representation's resolution diminishes, leading to an inevitable loss of information. As the exponent surpasses $1$, the vocabulary size becomes insufficient to effectively represent all underlying concepts. Consequently, communication increasingly relies on combinations of existing words to convey a broader conceptual space, which reduces efficiency.
        
        To explore this phenomenon, we analyze how the number of word combinations scales with corpus size $t$. Let $D_1(t) \sim t^{\beta_1}$ represent the number of distinct single words, $D_2(t) \sim t^{\beta_2}$ the number of distinct 2-grams, $D_3(t) \sim t^{\beta_3}$ the number of distinct 3-grams, and so forth. These relationships define higher-order Heaps' Laws, describing the growth of $n$-grams over time~\cite{di2025dynamics}. Each $n$-gram corresponds to a distinct concept that requires representation within the communication system.

        Constraints among higher-order Heaps' exponents $\beta_1, \beta_2, \beta_3, \dots$ naturally emerge. For example, constructing $D_2(t)$ distinct 2-grams requires at least $D_1(t) \sim \sqrt{D_2(t)}$ words, implying $\beta_1 \geq \beta_2 / 2$. Similarly, forming $D_3(t)$ distinct 3-grams requires $D_1(t) \sim D_3(t)^{1/3}$, leading to $\beta_1 \geq \beta_3 / 3$, and so forth. Approximating $D_2(t)$ as it grows linearly with corpus size (Fig.~\ref{fig:fig2}), with $\beta_2 \approx 1$, this constraint is therefore reflected in the Heaps' law of 1-grams, imposing that the vocabulary must grow at least as $D_1(t) \sim t^{0.5}$. This constraint may explain why Heaps' exponents below $0.5$ are rarely observed in natural language corpora, and it reflects the corresponding rarity of Zipf exponents exceeding $\alpha \approx 2$.
        
        The same constraint can be interpreted from a network perspective. Here, $D_1(t)$ represents the number of nodes, while $D_2(t)$ corresponds to the number of links in the network. The maximum number of possible links is $D_2^{\mathrm{max}} = D_1(t)(D_1(t)-1)/2 \sim D_1(t)^2$, which again gives the relation $\beta_1 \geq \beta_2 / 2$. If this constraint were violated, the network would eventually lack a sufficient number of nodes to sustain the creation of new links as it grows.

        At least for corpora of moderate size (not exceeding $t \sim 10^9$ words), the approximation $\beta_2 \approx 1$ ensures that a network representation of language comprehensively captures its statistical properties. From the fact that $D_1(t) \leq D_2(t) \leq D_3(t) \leq \cdots\leq D_\infty(t)$ we deduce that $\beta_1 \leq \beta_2 \leq \beta_3 \leq ...\leq 1$. Since $\beta_2 \approx 1$, it follows that $\beta_n \approx 1$ for all higher orders $n>2$. This exhausts the statistical description of $n$-grams, making it unnecessary to explicitly model higher-order networks, as their dynamics are inherently encoded in the observed relations. Although higher-order interactions exist, in our approximation, their principal statistical properties can be derived from the pairwise structure.
    
\section{Scale-free networks}
\label{app:B}

    \subsection{General properties} 
    
        Scale-free networks~\cite{barabasi1999emergence, barabasi2003scale} are networks whose node degree distribution follows a power law:
        \begin{equation}
            p(k) = C k^{-\gamma}
            \label{eq:power_law_degree}
        \end{equation}
        where $C$ is a normalization constant. Scale-free properties are ubiquitous in real-world networks, from the World Wide Web~\cite{adamic2000power}, to protein networks~\cite{wuchty2001scale, benz2008discovery}, and the Word co-occurrence Network~\cite{cancho2001small, kapustin2007vertex}, with degree exponents $\gamma$ typically ranging between 2 and 3.

        The main difference between random~\cite{erdos1960evolution} and scale-free networks is observed in the tail of the distribution. For example, we can compute the size of the largest hub $k_{\max}$, representing the degree of the most connected node in the network. This can be estimated by considering that the probability of observing the largest hub must be $1$ over the number of nodes $D$ in the network since there is at most one node with degree $k_{\max}$. This is expressed as:
        \begin{equation}
            \int_{k_{\max}}^{\infty} p(k) dk = \dfrac{1}{D}.
            \label{eq:kmax}
        \end{equation}
        For a random graph~\cite{erdos1960evolution} with degree distribution $p(k) = C e^{-\lambda k}$, Eq.~(\ref{eq:kmax}) gives $k_{\max} \sim \log D$. This indicates that the size of the largest hub scales logarithmically with the system order $D$. In contrast, for a scale-free network with degree distribution given by Eq.~(\ref{eq:power_law_degree}), we have:
        \begin{equation*}
            k_{\max} \sim D^{\frac{1}{\gamma - 1}}.
        \end{equation*}
        In this case, the size of the largest hub grows much more rapidly as the network size increases. In scale-free networks there can be several orders of magnitude between the smallest node, typically with degree 1, and the largest hub.
        
        The $n-$th moment of the degree distribution $p(k)$ is computed as:
        \begin{equation*}
            \langle k^n \rangle = \int_1^{k_{\max}} k^n p(k) dk = C \dfrac{k_{\max}^{n-\gamma+1}-1}{n-\gamma+1}
        \end{equation*}
        When $\gamma<3$, all moments of the degree distribution $p(k)$, except for the first, diverge with the system size. The divergence of $\langle k^2 \rangle$ indicates that fluctuations around the average degree can become arbitrarily large; if a node is randomly selected, its degree can be either small or diverge with the system order $D$. In this sense, these networks lack a characteristic scale, unlike random poissonian networks.
        
        Scale-free networks with $\gamma > 2$ exhibit small-world properties, providing high navigability. The diameter of such graphs, defined as the maximal distance between two nodes, is smaller than that of random graphs. In random graphs, the diameter scales as $d \sim \log D$. For scale-free networks, the diameter depends on the degree exponent $\gamma$, and the relations are~\cite{barabasi2003scale}:
        \begin{equation*}
        d = 
            \begin{cases}
                \log D \qquad \text{for } \gamma > 3 \\
                \dfrac{\log D}{\log \log D} \qquad \text{for } \gamma = 3 \\
                \log \log D \qquad \text{for } 2 < \gamma < 3
            \end{cases}
        \end{equation*}
        In particular, for $\gamma$ between 2 and 3, the structure provides optimal navigability, with a diameter much smaller than that of a random graph.
    
    \subsection{Anomalous scale-free networks} 

        \begin{table}[t]
            \centering
            \begin{tabular}{|c|c|c|c|c|c|c|}
                \hline
                & $k(s)\sim s^\xi$ & $k(R)\sim R^{-\eta}$ & $s(R)\sim R^{-\alpha}$ & $p(k)\sim k^{-\gamma}$ & $p(s)\sim s^{-\delta}$ & $D(t)\sim t^\beta$\\
                \hline
                 Kernel &  $k\sim s^{2/3}$ & $k(R)\sim R^{-2/3}$ & $s(R)\sim R^{-1}$ & $p(k)\sim k^{-5/2}$ & $p(s)\sim s^{-2}$ & $D(t)\sim t$\\
                 \hline
                 Periphery & $k\sim s$     & $k(R)\sim R^{-2}$   & $s(R)\sim R^{-2}$ & $p(k)\sim k^{-3/2}$ & $p(s)\sim s^{-3/2}$ & $D(t)\sim t^{1/2}$\\
                 \hline
            \end{tabular}
            \vspace{2mm}
            \caption{\textbf{Summary of the scaling exponents in the Word Co-occurrence Network.} $k(s)$ represents the relationship between degree $k$ and strength $s$~\cite{cancho2001small}; $k(R)$ and $s(R)$ denote the degree and strength rank-size distributions, respectively; $p(k)$ and $p(s)$ correspond to the probability distributions of degree and strength, respectively; and $D(t)$ describes the scaling of vocabulary size with corpus length. The scaling relations between these exponents are reported in Eq.~(\ref{eq:scaling}).}
            \label{tab:exponents}
        \end{table}

        When $\gamma < 2$, the network is in the so-called anomalous regime~\cite{albert2002statistical}. The first anomaly is that the size of the largest hub, which should scale as $k_{\max} \sim D^{\frac{1}{\gamma-1}}$, grows faster than the system order $D$. The average degree $\langle k \rangle$ also diverges. Since no node can have a degree larger than the total number of nodes, macroscopic networks with this property cannot exist.
        
        Such networks can be only observed if they exhibit a cutoff in the degree distribution for large degrees. This is the case of real-world networks with $\gamma < 2$~\cite{newman2003mixing, jovanovic2001modeling} and of generative models that produce such networks~\cite{seyed2006scale, barabasi2001deterministic, schneider2011scale}. The cutoff acts as a reservoir of nodes from which the anomalous structure with $\gamma < 2$ can draw connections. In fact, the divergence of the average degree indicates that these networks display far more links than typical scale-free networks with $\gamma > 2$. 
        
        These anomalies arise in systems that tend to create more connections than new nodes, as in the case of Gnutella peer-to-peer networks and software package networks mentioned in the main text. This process also generates the anomalous region in the Word Co-occurrence Network, where the tendency to create syntactic connections overcomes the production of new words. Interestingly, the number of links $|E| = D \langle k \rangle/2$ reaches its maximum scaling, $|E| \sim D^2$, when $\gamma = 1.5$. This is the situation observed in language networks. The value $\gamma = 1.5$ can be interpreted as another fundamental limit, beyond which it becomes challenging for networks to sustain smaller exponents. In real-world and generative networks, values of $\gamma$ smaller than $1.5$ are rarely observed. 
        
        These networks also show reduced navigability compared to those with $\gamma > 2$. In fact, their diameter scales as $\log D$, similar to random networks, due to the large number of nodes with degrees $k = 1$ and $2$, which create long chains connecting to the core~\cite{seyed2006scale}. On the other hand, their clustering coefficient does not vanish in large systems, unlike in networks with $\gamma > 2$, where $C \sim D^{2-\gamma}$. The high navigability of the kernel, whose connections can also be accessed by nodes outside it, combined with the high clustering coefficient of most nodes, produces a typical small-world structure~\cite{watts1998collective}. The Word Co-occurrence Network exhibits a very small diameter (around $d \sim 2-3$), along with a much larger clustering coefficient ($C = 0.687$) compared to random networks ($C \sim 10^{-4}$)~\cite{cancho2001small}. 

        A summary of the empirical scaling exponents observed in the Word Co-occurrence Network is given in Table~\ref{tab:exponents}. These exponents are interrelated through a set of scaling relations, which can be derived through straightforward algebraic manipulations:
        \begin{equation}
            \beta=\frac{1}{\alpha}, ~~ \alpha = \frac{1}{\delta - 1}, ~~ \gamma= 1+\frac{\delta-1}{\xi}, ~~ \eta = \frac{1}{\gamma-1}.
            \label{eq:scaling}
        \end{equation}

\section{Details on the model}
\label{app:C}

    \subsection{The network generative model} 

        \begin{figure*}[t]
        \centering
        \includegraphics[width=\linewidth]{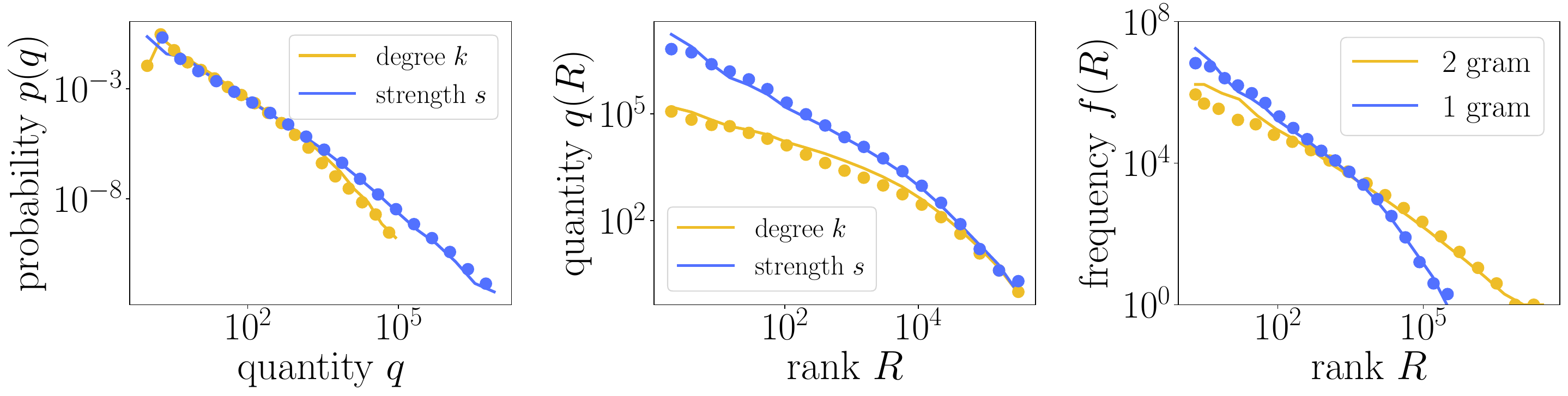}
        \caption{\textbf{Results of numerical simulations of the generative model.}  
        \textbf{Left:} Degree and strength distributions of the network generated by the model. Solid lines represent the simulation results, while points correspond to empirical data from the English language. The simulated network closely reproduces the empirical distribution, showing a kernel with $p(k) \sim k^{-1.5}$ and $p(s) \sim s^{-1.5}$, and a periphery with $p(k) \sim k^{-2.5}$ and $p(s) \sim s^{-2}$. \textbf{Center:} Degree and strength rank distributions. Simulations accurately follow the empirical trends, with $k(R) \sim R^{-0.66}$ and $s(R) \sim R^{-1}$ in the kernel, and $k(R) \sim R^{-2}$ and $s(R) \sim R^{-2}$ in the periphery. \textbf{Right:} Frequency-rank distributions of 1-grams and 2-grams from simulations. As observed in empirical data, the 2-gram distribution follows a power law, $f(R) \sim R^{-1}$. For 1-grams, the simulations reproduce the characteristic double-regime behavior: in the kernel lexicon, $f(R) \sim R^{-1}$, while in the periphery, $f(R) \sim R^{-2}$. All simulations were performed with $t=10^8$ iterations, $D_0 = 8000$ and $n_0 = 1$. Empirical data are retrieved from the Project Gutenberg corpus~\cite{gutenberg}, using a sequence of $10^8$ words.}

        \label{fig:fig6}
        \end{figure*}

        In this section we describe in detail the network generative model introduced in the main text.
        The generative model we propose to reproduce the properties of the Word Co-occurrence Network is inspired by the Urn Model with Triggering (UMT)~\cite{tria2014dynamics}, specifically its time-dependent version~\cite{bellina2024time, pemantle1990time}. This class of models is based on the Theory of the Adjacent Possible~\cite{kauffman1996investigations}, which posits that every time a new element is discovered, the space of possibilities expands. In other words, when a novelty is introduced, it opens up new possibilities of development. 
        
        In our model, the \textit{adjacent possible} consists of nodes that have not yet been visited. Nodes represent words, while edges correspond to their syntactic connections. Discovering a node for the first time is referred to as a ``novelty''. The model enlarges the adjacent possible space each time a novelty is encountered. Conversely, nodes that have already been explored are referred to as the \textit{actual space}, or \textit{actual network}. Specifically, the dynamics begin with a network composed of a certain number of nodes $n_0$\footnote{We fix $n_0=1$ all over this study. Any finite value of $n_0$ does not change the asymptotic results.}. These nodes are disconnected until they are visited for the first time. Nodes are visited with a probability proportional to their strength, and a link is created between any two nodes that are visited consecutively. If the link was already visited, its weight is incremented by one, simulating a reinforcement mechanism. This setup mirrors how the Word Co-occurrence Network is constructed, with the strength of a node increasing by 1 each time it is visited. In this sense, the model is also inspired by the class of edge-reinforced random walks~\cite{sabot2011edge}, particularly by the version incorporating triggering mechanisms~\cite{di2025dynamics}.

        The evolution of this class of models is typically expressed in terms of the frequency of occurrence of nodes in the network. Following the ideas of urn models, each time an element (word) $i$ is selected, it is reinforced by adding another element of the same type. In the network interpretation, this corresponds to increasing its weight by one and creating a link with the previously extracted word. Conversely, each time a novelty occurs—i.e., an element is extracted for the first time—the adjacent possible space expands by one element. Consequently, the size of the adjacent possible is equal to the number of novelties experienced, denoted as $D(t)$. In this sense, the model is equivalent to an Urn Model with Triggering (UMT) with parameters $\rho = \nu = 1$.

        The total number of elements in the urn is given by the initial number of elements $n_0$, plus the size of the actual space—which is equal to the total number of iterations $t$, since it increases by one at each extraction—plus the size of the adjacent possible, $D(t)$. Each element in the actual space is selected with a probability proportional to its frequency of occurrence in the sequence. Specifically, the probability is determined by the number of times element $i$ has been extracted, $f_i$, which corresponds to the number of elements of type $i$ in the urn. Conversely, each element in the adjacent possible appears only once. Thus, the frequency evolution follows:
        \begin{equation}
            \dfrac{df_i}{dt} = 
            \begin{cases}
                \dfrac{f_i}{n_0 + t + D(t)} \quad &\text{(actual network)}\\
                \dfrac{1}{n_0 + t + D(t)} \quad &\text{(adjacent possible)}
            \end{cases}
            \label{eq:freq_evolution}
        \end{equation}

        From a network perspective, this framework represents a system composed of two parts: the actual network and the adjacent possible. At each time step $t$, the probability of selecting a node from the actual network and the probability of introducing a new node from the adjacent possible are, respectively:
        \begin{equation}
            p_{\text{act}}(t) = \dfrac{t}{n_0 + t + D(t)}, \qquad \qquad p_{\text{adj}}(t) =\dfrac{n_0 + D(t)}{n_0 + t + D(t)}
            \label{eq:prob_adj}
        \end{equation}
        Thus, the probability of selecting a node from the actual space is proportional to the sum of links' weights in the actual network, given by $\sum_{i < j} w_{ij} = t$, since each at each extraction a link's weight $w_{ij}$ is increased by one. Conversely, the probability of extracting a node from the adjacent possible is proportional to its size, $n_0 + D(t)$.

        Within the actual space, the probability of selecting word (node) $i$ is proportional to its frequency and, consequently, to its strength:
        \begin{equation}
            \text{prob}_i(t) = \dfrac{f_i(t)}{\sum_{j=1}^D f_j(t)} = \dfrac{s_i(t)}{\sum_{j=1}^D s_j(t)}
            \label{eq:prob_strength}
        \end{equation}
        since, for nodes in the actual network, it simply holds that $s_i = 2 f_i$. This is because each time a node is extracted, it forms a link with both the preceding and following words, as in the Word Co-occurrence Network (WCN). This process shares similarities with the Barabási-Albert preferential attachment model, although the selection probability depends on strength rather than degree. However, at least in the initial stage, where $s_i \sim k_i$, the two processes exhibit strong similarities. Consequently, our model produces a scale-free network with a degree exponent between 2 and 3~\cite{barabasi1999emergence}.

        Conversely, when extracting from the adjacent possible, the selection is purely random among the $D(t)$ available elements. Each time a novelty is introduced, the adjacent possible grows by one node. The newly added node remains disconnected until it is visited for the first time, at which point it connects to the previously observed word, following the structure of a Word Co-occurrence Network.

        From the probability of extracting a node from the adjacent possible in Eq.~\ref{eq:prob_adj}, and thus increasing by one the vocabulary size $D(t)$, we can derive the following dynamical equation:
        \begin{equation}
            \dfrac{dD(t)}{dt} = \dfrac{n_0 + D(t)}{n_0 + D(t) + t} \approx \dfrac{D(t)}{D(t) + t}
            \label{eq:UMT}
        \end{equation}
        where the term $n_0$ can be neglected, as it remains constant over time.
        
        The final component of our model is a mechanism that reduces the probability of exploring the Adjacent Possible. This is achieved by introducing a time-dependent probability $p(t)$, which increases over time and reflects the growing tendency to explore syntactic connections between existing words rather than adding new terms to the vocabulary. This probability depends directly on the number of available syntactic connections, which, in network terms, corresponds to the number of edges. Therefore, we assume that $p(t)$ scales with the square of the number of nodes, $D(t)^2$. An illustrative description of the model is provided in Fig.~\ref{fig:fig4}.
        
        As discussed in the main text, $p(t)$ must saturate at $1/2$ due to structural limitations in the network. This preserves the Heaps' exponent from decreasing below $0.5$, maintaining the ability to explore new syntactic connections at any time. This reasoning is based on the approximation $D_2(t) \approx t$, which holds well even for large corpora (Fig.~\ref{fig:fig2}). The specific form of $p(t)$ we adopt is a quadratic sigmoid-like function:
        \begin{equation}
            p(t) = \dfrac{1}{2} \dfrac{\frac{D(t)^2}{D_0^2}}{1 + \frac{D(t)^2}{D_0^2}}
        \end{equation}
        This function ensures that $p(t)$ starts close to 0 for small $t$ and approaches $1/2$ for large $t$, particularly when $t \gg D_0$. The parameter $D_0$ separates the optimal regime, where exploration is efficient and Zipf's Law is observed, from the suboptimal regime where exploration becomes less efficient. $D_0$ can be interpreted as the size of the kernel lexicon, representing the number of words for which communication is fully optimized. Once the dictionary exceeds this limit, word production efficiency decreases.
        
\begin{figure*}[t]
        \centering
        \includegraphics[width=\linewidth]{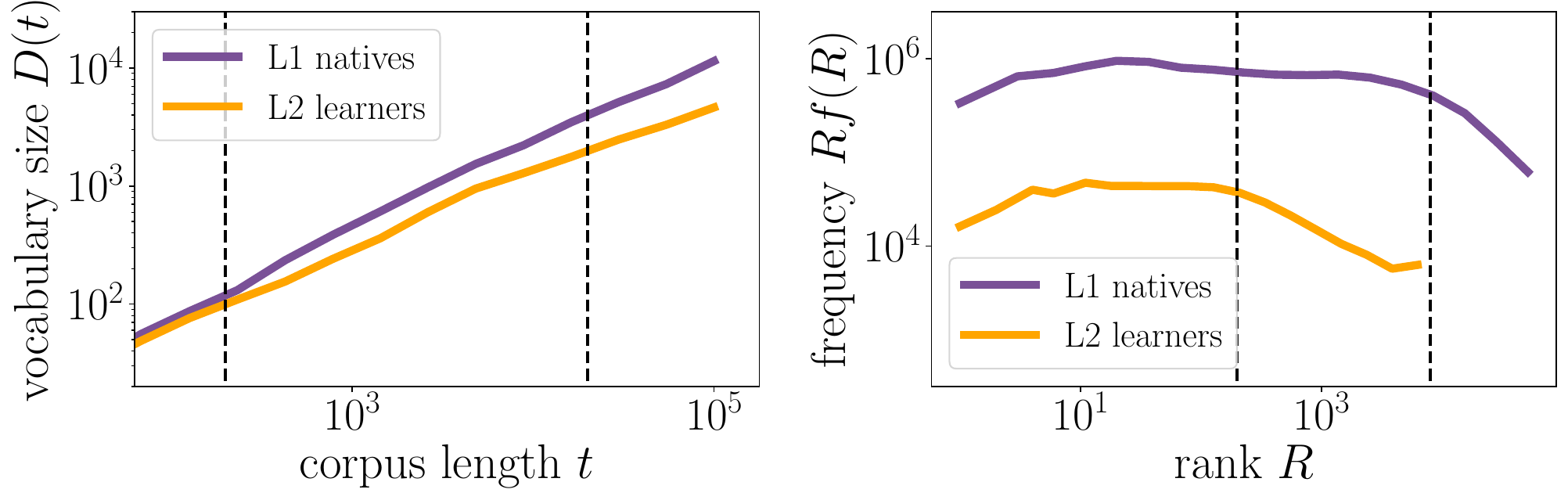}
        \caption{\textbf{Comparison between L1 and L2 corpus. Left:} Size of the dictionary $D(t)$ as a function of time $t$ for learners corpus (L2) and natives corpus (L1). For corpus of same length $t$ (set to $10^5$ words in this figure), the L1 corpus displays a larger dictionary with respect to L2. Native speaker do not only use a larger array of words, but they also use it more efficiently. This can be observed in the fact that the Heaps' exponent's change from $\beta=1$ to $\beta=0.5$ happens before in the L2 corpus with respect to L1, signalling a smaller size of the kernel lexicon. Dashed black lines indicate the point where the regime shift occurs. \textbf{Right:} Frequency rank distribution for L1 and L2 corpus. We multiplied the frequency $f(R)$ for the rank itself $R$ to highlight the change of the exponent from $1$ to $2$, which signals the size of the kernel lexicon. In the L1 corpus, $D_0 \sim 8000$, while for the L2 corpus only $D_0 = 200$. The values of $D_0$ are highlighted by the dashed black lines. This demonstrate the reduced capabilities of learners in producing an efficient communication, revealed in a smaller number of optimized words. The L2 corpus used for this figure comprises approximately $3.7 \cdot 10^5$ words. To better illustrate the two frequency-rank distribution regimes, we consider a larger L1 corpus, totaling $10^7$ words.}
        \label{fig:fig3}
        \end{figure*}

        As noted in Section~\ref{sec:model}, the size of the kernel lexicon can vary between different types of speakers, reflecting their cognitive and linguistic abilities to optimize communication. For instance, as shown in Fig.~\ref{fig:fig3}, learners of a language have a smaller $D_0$ compared to native speakers, indicating their reduced capacity to combine and produce words.
        
        The evolution of the number of nodes $D(t)$ changes with the introduction of $p(t)$, and becomes:
        \begin{equation}
            \dfrac{dD(t)}{dt} \approx (1 - p(t))\dfrac{D(t)}{D(t) + t}
        \end{equation}
        This dynamics is formally equivalent to a time-dependent UMT with $\nu(t) = 1 - p(t)$~\cite{bellina2024time}. This model replicates the scaling exponents observed in human language. The analytical solution of this model is reported in the following section.
        
    \subsection{Analytical solution of the model}

        In this section, we provide the analytical solution to the model proposed in this manuscript.
        
        We start by considering the differential equation that models the growth of $D(t)$, given by:
        \begin{equation*}
            \frac{dD}{dt} \approx \left(1 - p(t)\right) \frac{D(t)}{t + D(t)},
        \end{equation*}
        where $p(t)$ represents a probability function that depends on $t$ through $D(t)$. Specifically, we have:
        \begin{equation*}
            p(t) = \frac{1}{2} \frac{D^2(t)/D_0^2}{1 + D^2(t)/D_0^2},
        \end{equation*}
        which leads to the following expression for $\frac{dD}{dt}$:
        \begin{equation*}
            \frac{dD}{dt} \approx \left(1 - \frac{1}{2} \frac{D^2(t)/D_0^2}{1 + D^2(t)/D_0^2}\right) \frac{D(t)}{t}
        \end{equation*}
        where we have neglected the term $D(t)$ at denominator since $D(t) < t$. By separation of variables, we can rewrite this equation as:
        \begin{equation*}
            \frac{dD}{D(t)\left(1 - \frac{1}{2} \frac{D^2(t)/D_0^2}{1 + D^2(t)/D_0^2}\right)} = \frac{dt}{t}.
        \end{equation*}
        Integrating the right hand side between $t_0 = 1$ and $t$, and the left hand side between $D(0) = 1$ and $D(t)$ we obtain:
        \begin{equation*}
            \log D(t) + \frac{1}{2} \log \left[D^2(t) + 2 D_0^2\right] - \frac{1}{2} \log (2 D_0^2) = \log t.
        \end{equation*}
        From this, we derive the implicit equation for $D(t)$:
        \begin{equation*}
            D(t) \sqrt{D^2(t) + 2 D_0^2} = \sqrt{2} D_0 t.
        \end{equation*}
        Solving for $D(t)$, we get:
        \begin{equation*}
            D(t) = \frac{1}{2}\sqrt{- 4 D_0^2 + 2\sqrt{8 D_0^2 t^2 + 4 D_0^4}} = D_0 \sqrt{-1 + \sqrt{1 + 2\frac{t^2}{D_0^2}}}.
        \end{equation*}
        In the limit $t \ll D_0$ (using $\sqrt{1 + x} \approx 1 + 0.5 x$ for small $x$):
        \begin{equation*}
            D(t) \approx D_0 \sqrt{1 - 1 + \frac{t^2}{D_0^2}} = t,
        \end{equation*}
        which shows linear growth in the initial phase. In the limit $t \gg D_0$:
        \begin{equation*}
            D(t) \approx D_0 \sqrt{-1 + \sqrt{2\frac{t^2}{D_0^2}}} \approx D_0 \sqrt{\sqrt{2}\frac{t}{D_0}} = \sqrt{\sqrt{2} D_0} t^{0.5},
        \end{equation*}
        indicating a sublinear growth with exponent $0.5$ in the long-term limit.
        
        We can now obtain the frequency distribution from the cumulative distribution using the relations from~\cite{loreto2016dynamics}:
        \begin{equation*}
            p(r > R) \approx \frac{D(t/R)}{D(t)} = \frac{\sqrt{-1 + \sqrt{1 + 2\dfrac{t^2}{R^2 D_0^2}}}}{\sqrt{-1 + \sqrt{1 + 2\dfrac{t^2}{D_0^2}}}}.
        \end{equation*}
        In the limit $R \ll D_0$, i.e. for words in the kernel lexicon:
        \begin{equation*}
            p(r > R) \approx \frac{t/R}{t} = R^{-1},
        \end{equation*}
        The exponent of the frequency rank distribution is the inverse of that of the cumulative, thus we recover Zipf's Law with exponent $1$. For words outside the kernel lexicon, when $R \gg D_0$:
        \begin{equation*}
            p(r > R) \approx \frac{\sqrt{\sqrt{2} D_0} (t/R)^{0.5}}{\sqrt{\sqrt{2} D_0} t^{0.5}} = R^{-0.5},
        \end{equation*}
        which implies an exponent of $2$ in the frequency-rank distribution.

        From a network perspective, we expect from Eq.~(\ref{eq:prob_strength}) a scale-free structure in the kernel lexicon, with a degree exponent between 2 and 3. We adopt an exponent of $2.5$, which matches real data observations in Fig.~\ref{fig:fig2}. Hence, $p(k) \sim k^{-2.5}$. In general, for a probability distribution with exponent $\beta$, the cumulative distribution has an exponent $\beta + 1$. The frequency-rank distribution, whose exponent is the inverse of this, has exponent $\alpha = 1 / (1 + \beta)$. Given $\beta = -2.5$, we compute the frequency-rank degree distribution as $k(R) \sim R^{-0.66}$, consistent with real data. Since $f(R) \sim s(R) \sim R^{-1}$, we recover the relation between strength and degree in the kernel lexicon, $k(s) \sim s^{0.66}$. This agrees with previous empirical observations~\cite{cancho2001small}. Finally, from $s(R) \sim R^{-1}$ ($\alpha = 1$) and $\beta = 1 + 1/\alpha$, we obtain the strength probability distribution $p(s) \sim s^{-2}$, also in agreement with data.

        For words outside the kernel lexicon, the model gives $D(t) \sim t^{0.5}$ and $f(R) \sim R^{-2}$. In this part of the network, which includes less frequent words discovered later, the strength scales approximately with the degree, $s \sim k$. This implies that in the tail of the frequency-rank distribution, both strength and degree have the same exponent, giving $k(R) \sim R^{-2}$, consistent with observations. This also results in the same expression for the probability distributions, $p(s) \sim s^{-1.5}$ and $p(k) \sim k^{-1.5}$, confirming the origin of the anomalous exponent observed outside the kernel lexicon.

        The agreement between model simulations and real data is shown in Fig.~\ref{fig:fig6}. 

    \subsection{The size of the kernel lexicon, comparison between L1 and L2 corpora and cross-language comparison}
    \label{app:C3}

        The physical parameter of the model, $D_0$, regulates the size of the kernel lexicon, representing the optimized portion of vocabulary that follows Zipf's Law. In the model interpretation, $D_0$ is directly linked to the cognitive and linguistic capabilities of the speaker: the greater the speaker's ability, the larger the optimized vocabulary. This conjecture is supported by studies comparing different corpora for various types of speakers. For instance, prior research has shown that schizophrenic patients exhibit larger Zipf's exponents in small corpora~\cite{zipf2013psycho, ferrer2005variation}. However, the frequency-rank exponent rarely falls below $2$, indicating that variations between corpora primarily affect the kernel lexicon size rather than the tail exponent itself.
        
        To validate this hypothesis, we analyzed corpora of native English speakers (L1) and learners of English (L2)~\cite{lozano2020designing}. It is reasonable to assume that L2 speakers exhibit less optimized language usage compared to L1 speakers. This difference should manifest in a smaller kernel lexicon size, i.e., a reduced $D_0$ when fitting the model. This expectation is confirmed, as illustrated in Fig.~\ref{fig:fig3}: the kernel lexicon size in the L1 corpus is $D_0 \sim 8000$, while in the L2 corpus it is only $D_0 \sim 200$. L2 speakers rely on a smaller optimized vocabulary and compensate by heavily using combinations of words to express complex ideas. Consequently, their word frequency distribution exhibits a larger tail with an exponent of approximately $2$.

        These considerations, however, require additional care when comparing across languages. Although in many languages the kernel lexicon of native speakers is around $D_0 \sim 8000$, in some cases it appears significantly larger. For instance, German exhibits $D_0 \sim 20000$. This would imply that German speakers are more than twice as “capable” as speakers of other languages, but such interpretation is misleading. As previously noted, German has a strong morphological tendency to form compound words~\cite{fleischer2012word}. Linguistic studies show that over 60\% of the German lexicon consists of compounds~\cite{harlass1974aktuellen}, compared to only about 30\% in English~\cite{mcgregor2010compound}. This morphological feature may account for the larger size of the German kernel lexicon, as also noted in~\cite{gerlach2013stochastic}.

        If the larger size of $D_0$ is merely an artifact of compounding, it does not reflect a real gain in communicative efficiency. In practice, the additional German words included in the kernel lexicon are often concatenations of other base words which are equivalent to multi-word expressions in other languages, as illustrated in Fig.~\ref{fig:fig1}. German can thus express the same content using fewer word tokens~\cite{berg2017compounding}. However, because it does not introduce genuinely shorter representations of meaning, the total number of letters required remains approximately the same. The reduced word count is offset by longer average word length. Any saving in overall text length is therefore minimal.

        By contrast, the L1/L2 comparison within the same language is conceptually different. Native speakers produce shorter texts in terms of word count compared to learners, due to their larger kernel lexicon $D_0$. In this case, a smaller number of words also results in a smaller number of letters, since both groups use the same lexicon and grammar. Therefore, a smaller $D_0$ directly reflects a reduced efficiency in language use.

\section{Data sources and other languages analysis}
\label{app:D}

        \begin{figure*}[t]
        \centering
        \includegraphics[width=\linewidth]{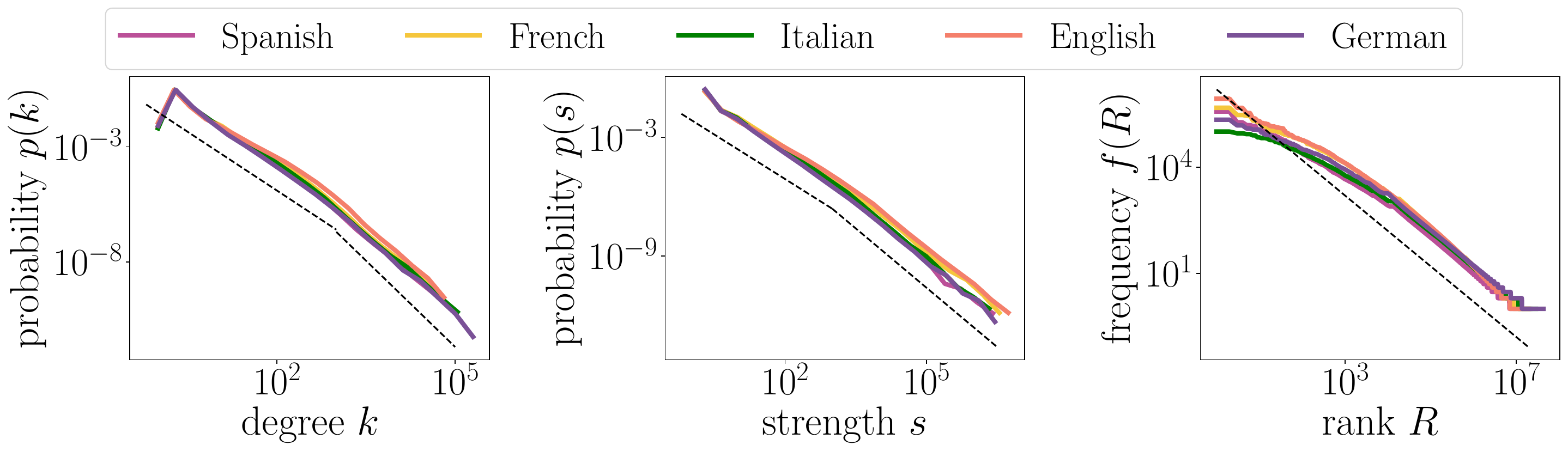}
        \caption{\textbf{Network properties across different languages.}  
        This figure summarizes the statistical properties of Word Co-occurrence Networks (WCNs) for multiple languages: English, German, French, Italian, and Spanish.  
        \textbf{Left:} Degree distribution of the WCNs. As observed in English, the network exhibits a core (kernel) following $p(k) \sim k^{-1.5}$, while the periphery follows $p(k) \sim k^{-2.5}$.  
        \textbf{Center:} Strength distribution of the WCNs. The kernel follows $p(s) \sim s^{-1.5}$, whereas the periphery follows $p(s) \sim s^{-2}$, consistent with the English case.  
        \textbf{Right:} Frequency-rank distribution of 2-grams. As in English, the distribution closely follows a power law with an exponent of 1, i.e., $f(R) \sim R^{-1}$.  
        These results demonstrate that the statistical properties discussed in the main text for English also hold similarly for other European languages.  
        Plots were generated using the Project Gutenberg dataset~\cite{gutenberg}, with a maximum $10^8$ words considered for each language.}
        \label{fig:all_languages}
        \end{figure*}

    The frequency-rank distribution of single words in Fig.~\ref{fig:fig1} (top left) was retrieved from the version 20200117 of the Google Ngrams dataset~\cite{michel2011quantitative}. The frequency-rank distribution of n-grams in Fig.~\ref{fig:fig1} (top right) was instead retrieved, due to computational contraints, from version 20120701, which is smaller than 20200117. We downloaded frequency data for 1- to 5-grams in English, Spanish, Italian, French, and German, filtering occurrences from 1950 onward. The Google Ngrams provides only word frequencies without access to the original text, though syntactic connections can be inferred from higher-order n-grams.

    The positions (rank) of the words highlighted in Fig~\ref{fig:fig1}, retrieved from the data, are: \textit{water}, 318; \textit{Wasser}, 305; \textit{bottle}, 5312; \textit{Flasche}, 6324; \textit{water bottle}, 322242; \textit{Wasserflasche}, 329067.
    
    To analyze the Word Co-occurrence Network (WCN), we used corpora from Project Gutenberg~\cite{gutenberg}, an online library containing over 70,000 free eBooks in multiple languages. Unlike Google Ngrams, this dataset includes complete texts, preserving actual syntactic structures and making it well-suited for network analysis.  

    We compiled five corpora from the Project Gutenberg database by selecting books in different languages. Specifically, we downloaded: 4615 books in English (total words: 284,470,403), 2231 books in German (total word count: 113,308,544), 2763 books in French (total word count: 151,368,265), 1093 books in Italian (total word count: 76,832,171), and 972 books in Spanish (total word count: 54,271,089). To estimate Heaps' Law, books are arranged in a random order. In Fig.~\ref{fig:all_languages}, we illustrate the statistical properties of word distributions for these five languages. The results confirm that the arguments presented in this manuscript, although primarily analyzed for English, also apply to other languages. In fact, other languages, at least European ones, exhibit statistical features that are approximately consistent with those observed in English.
    
    For the comparison between native speakers and learners, we used data from the COREFL database~\cite{lozano2020designing}. This contains transcribed speech and written texts produced by learners of English as a second/foreign language (L2), with Spanish or German as their first language (L1). The dataset amounts to a total of 2,342 participants and 530,392 words. Two ``control subcorpora'' are also included, consisting of texts in the participants' native languages and texts by English native speakers. For our analysis, we extracted only the corpus produced as a second language (L2), which resulted in a final word count of 372,167. 

\end{document}